\documentclass[iop,numberedappendix,apj]{emulateapj}
\usepackage{amsmath}
\usepackage{natbib}
\usepackage{epsfig}
\usepackage{xcolor}
\usepackage{graphicx}
\usepackage[colorlinks=true,linkcolor=blue,anchorcolor=blue,citecolor=blue,urlcolor=blue,draft=false]{hyperref}
\usepackage[caption = false]{subfig}

\newcommand{\msun}{\mbox{$M_{\sun}$}}
\newcommand{\lsun}{\mbox{$L_{\sun}$}}
\newcommand{\msunyr}{\mbox{$M_{\sun}~\textrm{yr}^{-1}$}}

\newcommand{\um}{\mbox{\rm $\mu$m}}

\newcommand{\spitzer}{\mbox {\it Spitzer}}
\newcommand{\galex}{\mbox {\it GALEX}}

\newcommand{\wise}{\mbox {\it WISE}}
\newcommand{\herschel}{\mbox {\it Herschel}}

\newcommand \acorr {$a_{\rm corr}$}

\begin{document}

\title{On the Spatially Resolved Star Formation History in M51 I: Hybrid UV+IR Star Formation Laws and IR Emission from Dust Heated by Old Stars}

\author{R. T. Eufrasio,\altaffilmark{1\dag}
        B. D. Lehmer,\altaffilmark{1}
        A. Zezas,\altaffilmark{2,3}
        E. Dwek,\altaffilmark{4}
        R. G. Arendt,\altaffilmark{4,5}
        A. Basu-Zych,\altaffilmark{4,5}
        T. Wiklind,\altaffilmark{6}
        M. Yukita,\altaffilmark{4,7}
        T. Fragos,\altaffilmark{8}
        A. E. Hornschemeier,\altaffilmark{4}
        L. Markwardt,\altaffilmark{1}
        A. Ptak,\altaffilmark{4} \&
        P. Tzanavaris\altaffilmark{4}
       }
\altaffiltext{1}{Department of Physics, University of Arkansas, 226 Physics Building, 825 West Dickson Street, Fayetteville, AR 72701, USA}
\altaffiltext{2}{Physics Department, University of Crete, Heraklion, Greece}
\altaffiltext{3}{Harvard-Smithsonian Center for Astrophysics, 60 Garden Street, Cambridge, MA 02138, USA}
\altaffiltext{4}{NASA Goddard Space Flight Center, Greenbelt, MD 20771, USA}
\altaffiltext{5}{Center for Space Science and Technology, University of Maryland Baltimore County, 1000 Hilltop Circle, Baltimore, MD 21250, USA}
\altaffiltext{6}{Physics Department, The Catholic University of America, Washington, DC 20064, USA}
\altaffiltext{7}{The Johns Hopkins University, Homewood Campus, Baltimore, MD 21218, USA}
\altaffiltext{8}{Geneva Observatory, Geneva University, Chemin des Maillettes 51, 1290 Sauverny, Switzerland}
\altaffiltext{\dag}{eufrasio@uark.edu}

\begin{abstract}
We present {\sc Lightning}, a new spectral energy distribution (SED) fitting procedure, capable of quickly and reliably recovering star formation history (SFH) and extinction parameters. The SFH is modeled as discrete steps in time. In this work, we assumed lookback times of 0--10\,Myr, 10--100\,Myr, 0.1--1\,Gyr, 1--5\,Gyr, and 5--13.6\,Gyr. {\sc Lightning} consists of a fully vectorized inversion algorithm to determine SFH step intensities and combines this with a grid-based approach to determine three extinction parameters. We apply our procedure to the extensive FUV-to-FIR photometric data of M51, convolved to a common spatial resolution and pixel scale, and make the resulting maps publicly available.  We recover, for M51a, a peak star formation rate (SFR) between 0.1 and 5\,Gyr ago, with much lower star formation activity over the last 100\,Myr. For M51b, we find a declining SFR toward the present day. In the outskirt regions of M51a, which includes regions between M51a and M51b, we recover a SFR peak between 0.1 and 1\,Gyr ago, which corresponds to the effects of the interaction between M51a and M51b. We utilize our results to (1) illustrate how UV+IR hybrid SFR laws vary across M51, and (2) provide first-order estimates for how the IR luminosity per unit stellar mass varies as a function of the stellar age. From the latter result, we find that IR emission from dust heated by stars is not always associated with young stars, and that the IR emission from M51b is primarily powered by stars older than 5\,Gyr.
\end{abstract}


\section{Introduction}{\label{sec:intro}}
The UV--to--IR spectral energy distribution (SED) of a galaxy, or galactic region, provides useful information about the star-formation history (SFH), which is defined as the star-formation rate (SFR) as a function of time. As byproducts of the SFH, additional useful quantities like the current SFR and
stellar mass ($M_\star$) can be obtained. Obtaining this valuable information from an SED requires modeling the evolution of the stellar populations and their metallicities throughout a long-term SFH, as well as the absorption and emission from gas and dust surrounding the stars and along the line of sight (see \citealt{Walcher2011} for a review and {\ttfamily http://sedfitting.org}).

Given the many components required to model an SED, several simplifying assumptions are often invoked.  For example, single burst or exponentially decaying SFHs (with one or two components typically), as well as single metallicities and rigid attenuation curves, are often assumed. Publicly available
SED fitting codes have become numerous, each having certain strengths and weaknesses, including, e.g., the treatment of the evolutionary phases of various stars (e.g., the AGB phase) in the core stellar evolution code, modeling of the IR emission from dust, and flexibility in assumptions (see, e.g., Conroy~2013 for a review of the trade-offs).  The variability of available SED fitting codes is a product of the needs of the users who developed them, as well as the continuous advancement in computing capabilities and physical models.

In this paper, we present a new SED fitting procedure that is part of a
long-term effort designed to understand (1) how hybrid UV+IR star formation laws vary
throughout galaxies due to variations in local SFHs and spatial scales; (2) how
stellar populations of various ages contribute to the heating of dust in the
interstellar medium; and (3) how X-ray binary populations evolve with time with
respect to their parent stellar populations.  Addressing these scientific areas
requires deriving spatially resolved ``maps'' of SFH information in nearby
galaxies using UV--to--IR data, with supplemental high spatial resolution X-ray
data (e.g., from {\it Chandra}) to associate X-ray binaries with local stellar
populations.  To this extent, our SED fitting procedure must be both fast, as to
obtain SFHs and associated errors over many thousands of pixels, and it must be sufficiently accurate in terms of obtaining SFH parameters and uncertainties consistent with those generated by more detailed SED fitting codes that take longer to implement due to large computational intensity.

Searching parameter space for the best solutions and assigning uncertainties to all model-derived quantities is usually very computationally intensive, with methods ranging from inversions \citep{Heavens2000, CidFernandes2005}, grid based $\chi^2$ minimization (e.g., \citealt{MentuchCooper2012}; \citealt{Boquien2016}), and an increasing number of models relying on Markov Chain Monte Carlo (MCMC; e.g., \citealt{Leja2017}). Recently \cite{Iyer2017} introduced what they called the basis dense approach, a procedure testing many SFH functional forms in order to derive a smooth SFH, after many regularization conditions.  Here, we implement SFHs as pre-defined, discretized steps in time and utilize a likelihood maximization based on matrix inversions to optimize computational efficiency, similar to \cite{Dye2008}, but simpler and faster. Our approach, implemented in the SED fitting code {\sc Lightning}, is easily hundreds to thousands of times less computationally intensive than many other methods and leads to consistent
results. This is primarily due to the fact that all the parameters describing our SFHs are determined by a series of matrix inversions, which usually requires much less computation than other methods, especially when dealing with large dimensional parameter spaces describing the SFH. This speed improvement is paramount when fitting a large number of SEDs across galaxies on a pixel-by-pixel basis, as is required by our long-term scientific goals described above.

In the long term, we will apply our method widely; however, here we initially apply our method to data in the Whirlpool galaxy (M51a or NGC5194) and its companion (M51b or NGC5195). M51 is an ideal first test case due to the proximity of the system (8.58~Mpc, which corresponds to a physical scale of
41.6~pc~arcsec$^{-1}$; \citealt{McQuinn2016}), large angular size on the sky ($\approx$8\arcmin$\times$12\arcmin), the nearly face-on orientation of M51a, the extensive multiwavelength coverage in the public archives, and the availability of many comparison results from the literature to test our procedure. In particular, the relatively well-determined timescale for the
interaction between M51a and M51b ($\approx$350--500~Myr ago; e.g., \citealt{SL2000}; \citealt{Dobbs2010}; \citealt{MentuchCooper2012}) allows us to test whether this epoch is recovered from our derived SFHs.  This paper focuses on the UV--to--IR SED fitting procedure, and presents first results related to the variations in the star-formation laws across the M51 system and the heating of dust from stellar populations of various ages. In a companion paper, Lehmer et~al.\ (2017, submitted), we present results related to the evolution of the X-ray binary luminosity function with time, which makes use of the SFH maps derived in this paper. The products derived in this paper (e.g., maps of SFH, SFR, $M_\star$, and extinction parameters), as well as the IDL codes that are used here, are
provided publicly at the Astrophysics Source Code Library and at {\ttfamily https://lehmer.uark.edu/}.  

This paper is structured as follows: in \S\,2, we present the
dataset and introduce the regions into which we will subdivide M51; in
\S\,3, we describe our SED fitting model and apply it to M51
on a pixel-by-pixel basis, in \S\,4, we compare our derived parameters with those presented in the literature. In \S\,5 provide first results on the spatial variations in the UV + IR star formation law and dust heating from stellar populations of different ages; finally, 
in \S\,6 we summarize our results.

\section{Data}{\label{sec:data}}

\subsection{Broadband Photometry}{\label{sec:photometry}}

A total of 18 bands from the FUV to the FIR were used in this work (see Table~\ref{tab:dataset}). Central wavelengths, FWHM spatial resolutions, normalized Galactic extinction, and fractional calibration uncertainties are listed. We have combined archival data from several facilities, including \galex, SDSS, 2MASS, \spitzer, and \herschel. All these data are publicly available and they were downloaded from their respective archives.

\subsection{Foreground Stars Recognition and Masking}{\label{sec:photometry}}

Foreground stars were masked and replaced with the median local background from the FUV to the \spitzer\ 4.5\,\um\ band. For longer wavelengths, the contribution from any foreground star is negligible.

First we used the \textsc{find} procedure of the NASA Goddard IDL Library \citep{Landsman93} to search for point sources over the whole field of the 3.6\um\ \spitzer\ image. Then we used the \textsc{aper} routine also from the Goddard Library to performed aperture photometry in all detected point sources in each \spitzer\ IRAC band (3.6, 4.5, 5.8, and 8.0\um). We used apertures with diameters of 2 times the FWHM, with a sky background determined between radii of 3 and 5 times the FWHM, with the FWHM varying according to each band. Specifically, we used FWHM of 1.7\arcsec, 1.7\arcsec, 1.9\arcsec, and 2.0\arcsec for 3.6, 4.5, 5.8, and 8.0\,\um\ bands, respectively. 

Star-forming regions in M51 and foreground stars exhibit very different \spitzer\ IRAC SEDs. Star-forming region SEDs contain Polycyclic Aromatic Hydrocarbon (PAH) features, which enhance the fluxes in the 5.8 and 8.0\um\ IRAC bands, while foreground stars in the Galaxy show declining SEDs for increasing wavelengths, corresponding to the Rayleigh-Jeans tail of the stellar emission ($F_\nu\propto\lambda^{-2}$). We therefore recognize as foreground stars all point sources with ratios $F_\nu(5.8\um)/F_\nu(3.6\um)<1$ and $F_\nu(8.0\um)/F_\nu(3.6\um)<1$.

Many of these stars are detected in one band, but not in others. For instance, several of them are not detected in the \galex\ FUV image and masking these undetected stars could mask real star-forming regions. We therefore only mask a star if it is detected with at least a signal-to-noise of 5, with photometry performed in a circle of 2 times the FWHM diameter and background evaluated between 5 and 6 times the FWHM. The FWHM of each band is shown Table~\ref{tab:dataset}. We replaced pixels within a circle of 5 times the FWHM diameter centered on each star with the median background pixel around that star in all bands, with the exception of the \galex\ bands, where the masked circles had diameters of 3 times the FWHM. Note the masked areas are larger than the apertures used for detection of the cores of the stars.

We tested masking a different numbers of selected foreground stars, but we ended up selecting the 30 brightest stars in the field as a good trade-off between masking bright stars and not blocking star-forming regions.

\begin{deluxetable}{lcccc}
\tabletypesize{\footnotesize}
\tablewidth{0in}
\tablecaption{Multi-wavelength data set assembled for M51 \label{tab:dataset}}
\tablehead{Telescope/Band & $\lambda_0$\tablenotemark{a}& 
           FWHM\tablenotemark{b} & $f_{\rm Gal}({\lambda_0})$\tablenotemark{c} & $\sigma_{\rm cal}$\tablenotemark{d}}
\startdata
\galex\ FUV       &   0.152~\um & 4.3\arcsec  &  2.674  & $15\%$ \\
\galex\ NUV       &   0.226~\um & 5.3\arcsec  &  2.639  & $15\%$ \\
SDSS $u$          &   0.355~\um & 1.8\arcsec  &  1.547  &  $5\%$ \\
SDSS $g$          &   0.465~\um & 1.6\arcsec  &  1.211  &  $5\%$ \\
SDSS $r$          &   0.616~\um & 1.5\arcsec  &  0.832  &  $5\%$ \\
SDSS $i$          &   0.748~\um & 1.6\arcsec  &  0.621  &  $5\%$ \\
SDSS $z$          &   0.886~\um & 1.6\arcsec  &  0.463  &  $5\%$ \\
2MASS $J$         &   1.23~\um  & 3.0\arcsec  &  0.263  & $10\%$ \\
2MASS $H$         &   1.64~\um  & 3.0\arcsec  &  0.168  & $10\%$ \\
2MASS $K_{\rm S}$ &   2.16~\um  & 3.0\arcsec  &  0.105  & $10\%$ \\
\spitzer\ IRAC    &   3.6~\um   & 1.7\arcsec  &  0.046  & $ 5\%$ \\
\spitzer\ IRAC    &   4.5~\um   & 1.7\arcsec  &  0.029  & $ 5\%$ \\
\spitzer\ IRAC    &   5.8~\um   & 1.9\arcsec  & \nodata & $ 5\%$ \\
\spitzer\ IRAC    &   8.0~\um   & 2.0\arcsec  & \nodata & $ 5\%$ \\
\spitzer\ MIPS    &  24~\um     & 6.0\arcsec  & \nodata & $ 5\%$ \\
\herschel\ PACS   &  70~\um     &  5.2\arcsec & \nodata & $10\%$ \\
\herschel\ PACS   & 160~\um     & 12.0\arcsec & \nodata & $10\%$ \\
\herschel\ SPIRE  & 250~\um     & 18.0\arcsec & \nodata & $15\%$
\enddata
\tablenotetext{a}{Central wavelength ($\lambda_0$) of the filter, assuming a flat $F_\nu$ source, i.e., $\lambda_0 = \int_0^\infty\lambda R_k(\nu)\,d\nu / \int_0^\infty R_k(\nu)\,d\nu$. For the \spitzer\ and \herschel\ bands we display the nominal values, most commonly used.}
\tablenotetext{b}{FWHM spatial resolution of each map.}
\tablenotetext{c}{Galactic extinction curve normalized at the $V$ band, $f_{\rm Gal}(\lambda_0)=\tau_{\lambda_0}^{\rm Gal}/\tau_V^{\rm Gal}=A_{\lambda_0}^{\rm Gal}/A_V^{\rm Gal}$. $V$-band Galactic extinction of $A_V^{\rm Gal} = 0.095$.}
\tablenotetext{d}{Calibration uncertainty as a fraction of the intensity.}
\end{deluxetable}

\subsection{Convolution to a Common Spatial Resolution}{\label{sec:convolution}}
After masking foreground stars and replacing them with the local backgrounds, all maps were convolved to a common 25\arcsec\ FWHM spatial resolution and registered to a common astrometric frame with 10\arcsec\ pixels. We transform each native point spread function (PSF) to a final PSF of our choice, using the properties of Fourier transforms of convolutions. The procedure is described in \cite{Eufrasio2015} and \cite{Straughn2015} and it is made available to the public through the Astrophysics Source Code Library.

The worst spatial resolution image of our data set is the \herschel\ SPIRE 250\,\um, which has a FWHM resolution of approximately 18\arcsec. We have chosen 25\arcsec\ to err on the side of caution and guarantee our results do not depend on the uncertainties of the largest PSFs, mainly the 250\,\um\ and the \herschel\ PACS 160\,\um\ (12.0\arcsec). The approximate original FWHM spatial resolutions of all our images are tabulated in Table~\ref{tab:dataset}. We could in principle have done this analysis with better spatial resolution ($\sim$18\arcsec\ FWHM) and smaller pixels (6\arcsec$-$8\arcsec), but our spatial resolution of 25\arcsec\ allows us to include \herschel\ SPIRE 350\,\um\ in any subsequent study and a pixel size of 10\arcsec\ considerably decreases the number of SEDs to be fitted, making the process more tractable.

\begin{figure*}
  \singlespace
  \centering
  \includegraphics[width=.7\textwidth]{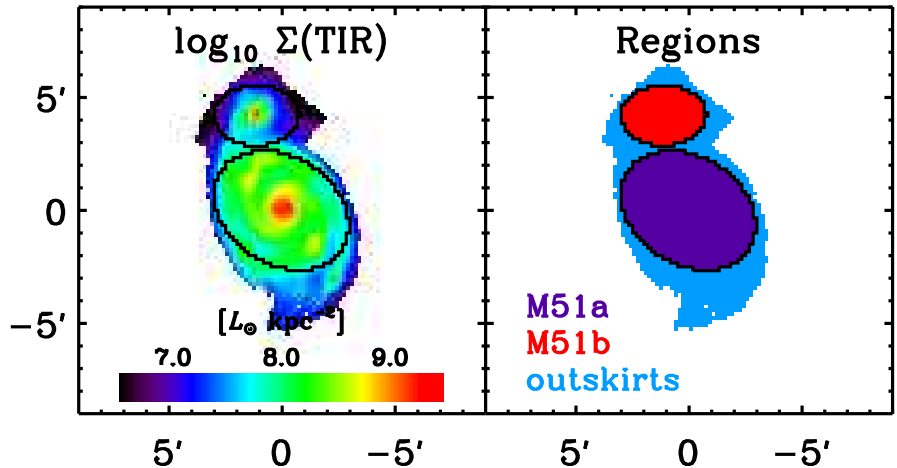}
  \vspace{-0.1in}
  \caption{Left panel shows the logarithm of the TIR surface density $\Sigma({\rm TIR})$, in \lsun~kpc$^{-2}$, obtained following Equation~\ref{eq:lir}. Right panel shows the pixels enclosed in the M51a and M51b ellipses (respectively in purple and red) and outside of them, i.e., in the outskirts region (in blue).}
  \label{fig:lir_map}
\end{figure*}

After matching spatial resolutions, we combined the \spitzer\ MIPS 24\,\um, the \herschel\ PACS 70\,\um\ and 160\,\um, and SPIRE 250\,\um\ maps in order to generate an IR map corresponding to the integrated TIR luminosity from dust. Our TIR luminosities correspond to the dust emission from 3 to 1100\,\um, based on the calibration from \cite{Galametz2013}, more precisely
\begin{equation}\label{eq:lir}
L_{\rm TIR} \equiv 2.127L_{24} + 0.702L_{70} + 0.974L_{160} + 0.382L_{250},
\end{equation}
where for each pixel of the image $L_{24}= \nu L_\nu(24~\um)$ and analogously for all bands. This calibration maps these four FIR bands into \cite{DL2007} models to recover appropriate TIR luminosities. The 1$\sigma$ statistical uncertainties for the coefficients in Equation~\ref{eq:lir} are 0.092, 0.024, 0.024, and 0.063, respectively. Since $L_{70}$ or $L_{160}$ are the larger of the four luminosities, the final TIR uncertainty is dominated by calibration uncertainties ($\sim$ 10\%; see Table~\ref{tab:dataset}) and not by these statistical uncertainties. The combined TIR surface density map is displayed in the left panel of Figure~\ref{fig:lir_map}.

\subsection{Background Subtraction}{\label{sec:background}}
The galaxies were masked to include all diffuse emission seen around the galaxies in all bands. Background values outside of the galaxy and their uncertainties were determined following the procedures of \cite{Eufrasio2014}. The final uncertainties include background subtraction and calibration uncertainties. Calibration uncertainties used for each band are listed in Table~\ref{tab:dataset}.

After background subtraction, we confined our analysis to pixels with signal-to-noise ratio greater than 4 in all bands. We tested different criteria, but this fully enclosed our pixels of interest. This left us with a total of 2043 pixels.

Determining dust emission parameters is beyond the scope of this paper and therefore we do not fit wavelengths longer than 4.5\,\um. Our final dataset therefore consists of 2043 generalized SEDs with 13 datapoints each (the first 12 broadband images listed in Table~\ref{tab:dataset} and the TIR luminosity).

\subsection{Regions}{\label{sec:regions}}
We defined larger regions to simplify some of our later analyses. These regions are shown in the right panel of Figure~\ref{fig:lir_map}. The largest region, corresponding to M51a, is shown in violet. In red, we display the pixels enclosed by the M51b ellipse. Both ellipses are used here as defined by \cite{MentuchCooper2012} and their parameters are given in Table~\ref{tab:regions}. In cyan, we show the pixels outside of both ellipses, which comprise the region we will refer hereafter as the ``outskirts". We also assess the properties of a region that includes all 2043 pixels combined.

\begin{deluxetable}{ccccc}
\tabletypesize{\footnotesize}
\tablewidth{0in}
\tablecaption{Region Characteristics}
\tablehead{ & M51a & M51b & outskirts & all }
\startdata
$\alpha_{\rm J2000}$ &    202.47065 &  202.49726 &  \nodata &  \nodata \\
$\delta_{\rm J2000}$ &     47.19517 &   47.26531 &  \nodata &  \nodata \\
$a$                  & 191.5\arcsec & 113\arcsec &  \nodata &  \nodata \\
$b$                  & 143.0\arcsec &  80\arcsec &  \nodata &  \nodata \\
PA                   &    57.5\degr &  90.0\degr &  \nodata &  \nodata \\
\#(pixels) & 862  & 284  & 897  & 2043 \\
Area [kpc$^2$] & 149.1 & 49.1 & 155.2 & 353.4
\enddata
\label{tab:regions}
\tablenotetext{}{Positions from \cite{MentuchCooper2012}. }
\end{deluxetable}

\begin{figure*}
  \singlespace
  \centering
  \includegraphics[width=0.7\textwidth]{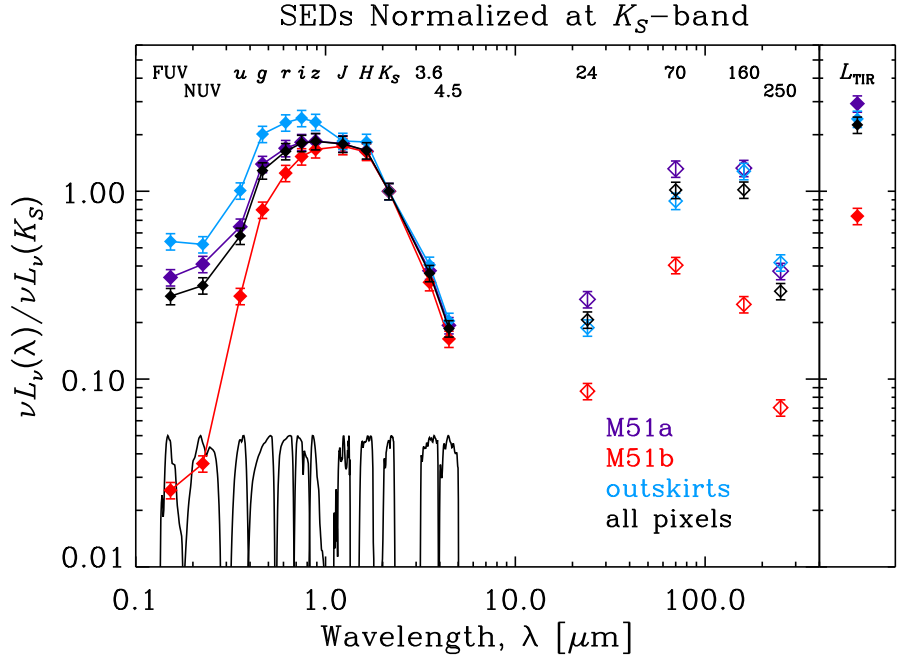}
  \vspace{-0.1in}
  \caption{SEDs normalized at the 2MASS $K_S$-band (i.e., 2.2\,\micron)for the large regions defined in Figure~\ref{fig:lir_map}. The color code is the same as in Figure~\ref{fig:lir_map}  Table~\ref{tab:regions} (i.e., violet, red, cyan, and black respectively correspond to M51a, M51b, outskirts, and all pixels combined). Uncertainties of 10\% were used just for plotting purposes, once the combined uncertainty is mostly coming from calibration uncertainties. Filled diamonds correspond to the 12 broadband filters and the TIR luminosity (shown in the side panel). Empty diamonds show the four FIR bands combined to produce the TIR map. Black curves correspond to the response functions of the broadband filters used in the SED fitting. Response function peaks were normalized to 0.05.}
  \label{fig:sed_norm}
\end{figure*}

In Figure~\ref{fig:sed_norm} we show the SEDs for all these regions, normalized at the $K_S$-band. The filled diamonds correspond to the 13 points of our SEDs and the empty diamonds show the four FIR bands combined to produce the TIR luminosity (displayed in the side panel). These SEDs are qualitatively quite different. M51b has the reddest stellar population and, at least apparently, the hottest dust emission in the FIR (from the highest ratio of 70 and 160\,\um\ luminosities). The outskirts is the bluest region in the UV/optical, indicating it was likely strongly affected by the interaction when star formation was triggered in this area. M51a has an intermediate SED between the previous two and the SED combining all pixels is the {\color{black}sum} of the three others.

\section{SED Modeling}\label{sec:SED_modeling}

\subsection{Step Star Formation History}{\label{sec:StepSFH}}

With our code {\sc Lightning}, the SFH of each pixel is modeled with $n$ steps in time, in the form of 
\begin{equation}
\psi(t) = \psi_i,\text{\quad for \quad}t_i < t < t_{i+1},
\end{equation}
where $t_i$ and $t_{i+1}$ represent the respective lower and upper boundaries to the $i$-th time step.

For the current work, we have chosen to model the SFH with five steps (i.e., $n=5$). The boundaries of the steps have lookback times of 0, 10\,Myr, 100\,Myr, 1\,Gyr, 5\,Gyr, and 13.6\,Gyr. Figure~{\ref{fig:steps_templates} shows the spectra corresponding to each of the five SFH steps, for $\psi_i=1$\,\msunyr. We have used the population synthesis code {\sc P\'egase} \citep{FRV1997,FRV1999}, with a \cite{Kroupa2001} IMF and a solar metallicity ($Z=0.02$).

\begin{figure}
  \singlespace
  \hspace{-.5in}
  \includegraphics[width=0.55\textwidth]{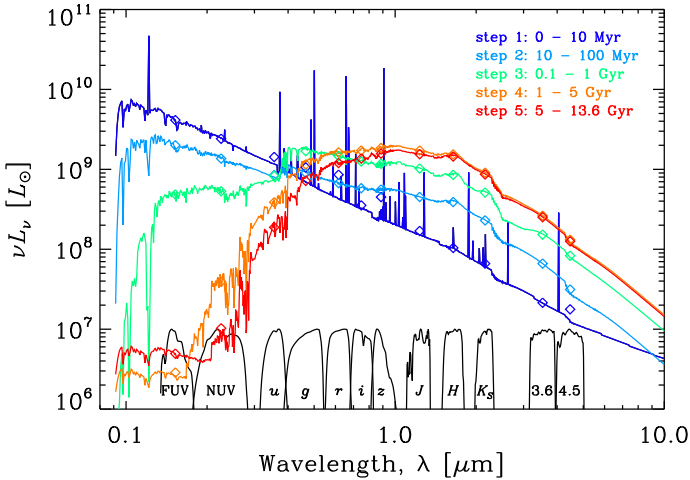}
  \vspace{-0.2in}
  \caption{Five spectral templates corresponding to the 5 steps of the SFHs considered in this paper. High-resolution {\sc P\'egase} spectra are shown in conjunction with the corresponding broadband SEDs. These are the intrinsic, unattenuated spectra, for a constant SFR of 1\,\msunyr. Only the first step (ages from 0 to 10 Myr) emit nebular continua and line emission. Table \ref{tab:sfh_facts} lists some properties associated with each of these spectral templates. Response functions of all filters are shown in black and their peaks were normalized to 10$^7$\,\lsun.}
  \label{fig:steps_templates}
\end{figure}

The first SFH step, from 0 to 10\,Myr, is chosen to model the youngest stellar population. This population is possibly still subjected to a higher birth-cloud attenuation and the only able to emit a substantial number of hydrogen-ionizing photons that quickly recombines and produces hydrogen recombination continua and lines. Its intensity $\psi_1$ is directly proportional to the total ionizing photon rate and the intrinsic H$\alpha$ line intensity.

The second SFH step, from 10 to 100\,Myr, models the stellar population that, combined with the first step, produces the bulk of the UV emission. Together with the first step it encompasses the typical timescale by which the current SFR is traced by the UV emission, but it still allows the average SFR averaged over the last 100\,Myr to be modeled by the two quantities $\psi_1$ and $\psi_2$.

The other three steps were chosen so we had five total SFH steps, as derived by \cite{Dye2008}, and similar bolometric luminosities for each of the three steps. \cite{Dye2008}, however, did not use UV data and did not impose energy balance between the absorbed UV/optical and dust-reprocessed IR, which considerably improves SED fits, as we will discuss later. \cite{Leja2017} used six SFH steps to model the integrated SED of the 129 nearby galaxies presented in \cite{Brown2014}. They were able to recover all six SFH steps relatively well from broadband SEDs alone. They have also observed remarkable agreement between the derived and observed H$\alpha$, even when the line emission is not used in the fits. Table~\ref{tab:sfh_facts} lists the stellar masses, bolometric luminosities, and the rate of hydrogen-ionizing photons $Q_0$ for each SFH step per 1\,\msunyr.
	
The extinction-free high-resolution model spectrum, $\tilde{L}_\nu(\nu)$, is therefore given by the sum of the spectra emitted from all steps of the SFH, $\tilde{L}_\nu^i(\nu)$, i.e.,
\begin{equation}
  \tilde{L}_\nu(\nu) = \sum\limits_{i=1}^n \psi_i \tilde{L}_\nu^{i}(\nu),
\end{equation}
where $\tilde{L}_\nu^{i}(\nu)$ is a specific luminosity per unit SFR, or the template spectrum arising for the $i$-th SFH step per $1$\,\msunyr, as depicted in Figure~\ref{fig:steps_templates}.

\begin{deluxetable}{ccccc}
\tabletypesize{\footnotesize}
\tablewidth{0in}
\tablecaption{SFH Steps for $\psi_i$ fixed at 1\,\msunyr \label{tab:sfh_facts}}
\tablehead{SFH Step  & Boundaries &   $M_\star$    & $L_{\rm bol}$ & $Q_0$ \\
            \#       &   [Gyr]    & [10$^9$\msun] & [10$^9$\lsun] & [10$^{53}$ph s$^{-1}$] }
\startdata
1 &    0 -- 0.01 & 0.0095 & 6.16 & 1.20  \\
2 & 0.01 -- 0.1  & 0.070  & 3.23 & 0.003 \\
3 &  0.1 -- 1    & 0.59   & 2.89 & 0     \\
4 &    1 -- 5    & 2.14   & 3.05 & 0     \\
5 &    5 -- 13.6 & 3.99   & 2.52 & 0
\enddata
\end{deluxetable}

\subsection{Dust Extinction Parameters}{\label{sec:extinction}}
The high-resolution model spectrum is attenuated by an intrinsic extinction curve and a fixed Galactic extinction curve. For the Galactic extinction, we use the \cite{F99} standard curve as coded in the \textsc{fm\_unred} script of the NASA Goddard IDL Library \citep{Landsman93}.

For the intrinsic attenuation, we consider a variable extinction curve composed of a diffuse attenuation component, common to steps of all ages,
\begin{align}\label{eq:tau_diff_lambda}
  \tau_{\rm DIFF}(\lambda)
  = \frac{\tau_{{\rm DIFF},V}}{4.05}\left[k^\prime(\lambda)+D(\lambda)\right]\left(\frac{\lambda}{\rm 5500\,\AA}\right)^\delta
\end{align}
$k^\prime(\lambda)$ is the \cite{Calzetti2000} extinction curve, as modified by \cite{Noll2009} with the addition of a bump and a $\delta$ parameter controlling the UV slope. The $D(\lambda)$ term adds a Drude profile representing the 2200\,\AA\ bump of the extinction curve, of the form
\begin{equation}
  D(\lambda) = \frac{E_b\left(\lambda\,\Delta\lambda\right)^2}
                    {\left(\lambda^2-\lambda_0^2\right)^2 + \left(\lambda\,\Delta\lambda\right)^2},
\end{equation}
where $\lambda_0=2175$\,\AA\ is the central wavelength of the bump and $\Delta\lambda=350$\,\AA\ its FWHM. In this work, we take the bump strength $E_b$ to be related to the slope $\delta$ as
\begin{equation}
  E_b = 0.85 - 1.9\times\delta,
\end{equation}
following the findings of \cite{KriekConroy2013}.

On top of the diffuse extinction, a birth-cloud attenuation is applied to the youngest SFH step, i.e., lookback times up to 10\,Myr. This extinction is given by 
\begin{equation}
  \tau_{\rm BC}(\lambda) = \tau_{{\rm BC},V}\left(\frac{\lambda}{\rm 0.55\,\um}\right)^{-1}.
\end{equation}

Thus, the attenuated, high resolution spectrum is
\begin{equation}
  L_\nu^{\rm mod}(\nu) = \sum\limits_{i=1}^n \psi_i L_\nu^{i}(\nu),
\end{equation}
with
$\displaystyle L_\nu^{i}(\nu) = \tilde{L}_\nu^{i}(\nu)~e^{-\tau_{\rm DIFF}(\nu)}e^{-\delta_{i1}\tau_{\rm BC}(\nu)}e^{-\tau_{\rm Gal}(\nu)}$, where $\delta_{i1}$ is the Kronecker delta, i.e., $\delta_{i1}$ equals unity for $i=1$ and 0 otherwise, not to be confused with the $\delta$ extinction parameter defined in Equation~\ref{eq:tau_diff_lambda}. Our model has therefore three free parameters that control extinction: $\tau_{{\rm DIFF},V}$, $\delta$, and  $\tau_{{\rm BC},V}$.

After extinction is applied, the high resolution attenuated spectrum is convolved with the broad or narrowband filters to produce the model SED that will be compared to the observations. The full SED model is then
\begin{equation}
  L_{\nu,k}^{\rm mod} = \sum\limits_{i=1}^n \psi_i L_{\nu,k}^{i}
                      =  \sum\limits_{i=1}^n \psi_i\displaystyle \int_0^\infty L_\nu^{i}(\nu)\,R_k(\nu)\,d\nu,
\end{equation}
with $R_k(\nu)$ being the response curve of the $k$-th filter, normalized to 1, i.e., $\int_0^\infty R_k(\nu)\,d\nu=1$.

The TIR dust emission provides a strong constraint on the attenuation, since the UV-optical attenuated light will be re-emitted in the IR by the absorbing dust particles. This paper aims at recovering SFHs and not the dust parameters, as done by \cite{MentuchCooper2012} and others, so we do not model the dust emission in detail. Instead, we have combined four FIR bands (24, 70, 160, and 250~\um) in a single total luminosity map, following Equation~\ref{eq:lir}.

In order to better constrain the attenuation parameters, we model the TIR luminosity as the total intrinsic (extinction-free) luminosity minus the attenuated luminosity. The model TIR luminosity $L_{\rm TIR}^i$ is separately computed for each SFH step, given the three extinction parameters, as
\begin{equation}\label{eq:ltir}
  L_{\rm TIR}^i = \int_0^\infty\left(\tilde{L}_\nu^i(\nu)-L_\nu^i(\nu)\right)d\nu
\end{equation}
and the model TIR luminosity is then
\begin{equation}
  L_{\rm TIR}^{\rm mod} = \sum_{i=1}^n\psi_i L_{\rm TIR}^i.
\end{equation}

The set of observables to be fitted	are therefore all the broadband luminosities as well as the TIR luminosity, i.e.,
\begin{equation}
  \left\{L_k^{\rm mod}\right\}\equiv\left\{L_{\nu,1}^{\rm mod},L_{\nu,2}^{\rm mod},\cdots,L_{\nu,m}^{\rm mod},L_{\rm TIR}^{\rm mod}\right\},
\end{equation}
so that, for any band $k$, 
\begin{equation}\label{eq:sed_model}
  L_k^{\rm mod} = \sum_{i=1}^{n} \psi_i L_k^i.
\end{equation}
$L_k^i$ is therefore the intensity of the $i$-th step in the $k$-th band, with $k$ varying from 1 to $m+1$, with $k=m+1$ representing the TIR luminosity of that given step per unit SFR, and a given set of the three extinction parameters.

\subsection{Likelihood Maximization via Matrix Inversion}{\label{sec:chi2minimization}}
For a given choice of extinction parameters, the model SED is a linear combination of the SEDs of the SFH steps, as in Equation~\ref{eq:sed_model}. This greatly simplifies $\chi^2$ minimization, as we show in the following lines, increasing computational speed for a large number of computations. We measure the goodness-of-fit as
\begin{align}\label{eq:sed_model_chisqr}
\chi^2 &= \sum_{k=1}^{m+1}\frac{\left(L_k^{\rm mod} - L_k^{\rm obs}\right)^2}
                              {\sigma_k^2} \nonumber\\
       &= \sum_{k=1}^{m+1}\frac{1}{\sigma_k^2}\left(\sum_{i=1}^n\psi_i L_k^{i} - L_k^{\rm obs}\right)^2.
\end{align}
Here $L_k^{\rm obs}$ are the observed luminosities and $\sigma_k$ are their corresponding $1\sigma$ uncertainties.

To minimize $\chi^2$, we set its partial derivatives with respect to each SFH step intensity $\psi_i$ equal to zero, i.e.,
\begin{equation}
  \frac{\partial\chi^2}{\partial\psi_i}
   = \sum_{k=1}^{m+1}\frac{2L_k^i}{\sigma_k^2}\left(\sum_{j=1}^n\psi_j L_k^{j} - L_k^{\rm obs}\right) = 0,
\end{equation}
which can be rearranged as
\begin{equation}
  \sum_{j=1}^n\left(\sum_{k=1}^{m+1}\frac{L_k^iL_k^{j}}{\sigma_k^2}\right)\psi_j
   = \left(\sum_{k=1}^{m+1}\frac{L_k^iL_k^{\rm obs}}{\sigma_k^2}\right)
\end{equation}
and simplified to
\begin{equation}\label{eq:matrix_elements}
  \sum_{j=1}^nA_{ij}\psi_j = B_i,
\end{equation}
where
\begin{equation}
A_{ij} \equiv \sum_{k=1}^{m+1}\frac{L_k^iL_k^{j}}{\sigma_k^2}
\end{equation}
is calculated from the population synthesis models alone and
\begin{equation}
B_i  \equiv \sum_{k=1}^{m+1}\frac{L_k^iL_k^{\rm obs}}{\sigma_k^2}
\end{equation}
is calculated from a combination of models and observed luminosities.

This procedure leads to a linear system of $n$ equations, that is, with Equations~\ref{eq:matrix_elements} representing as many equations as SFH steps. Hence the whole system of equations can be written in matrix form as
\begin{equation}
A \psi = B,
\end{equation}
where $A$ is a square ($n\times n$) symmetric matrix, with only positive elements, $\psi$ is a column vector with all the SFH coefficients $\psi_i$, and $B$ is column vector of dimension $n$ as well. $A$ is invertible if $\det(A)\neq0$ and in that case there is only a single solution for the system of equations. The solution for the SFH is thus given by
\begin{equation}\label{eq:firstinversion}
\psi = A^{-1}B.
\end{equation}

Since the coefficients $\psi_i$ represent the SFRs of the different previously defined SFH steps, they are necessarily non-negative real numbers. However, the numerical inversion in Equation~\ref{eq:firstinversion} does not require that $\psi_i\geq0$, for all steps $i$. The solution may initially contain one or more negative $\psi_i$ coefficients after the first matrix inversion.

To assure all SFH steps are non-negative, we have implemented an iterative algorithm that first sets to zero all coefficients that are found to be negative, next reduces the dimensionality of the problem, and finally solves Equation~\ref{eq:firstinversion} again for the other coefficients until they are all non-negative.

In order to determine the uncertainties in the SFH solutions, we simulate 400 perturbed SEDs and solve for the SFH each time. For each simulated perturbed SED, the fluxes of each band are drawn independently from Gaussian distributions set by the mean and standard deviation of the measured fluxes at each wavelength.

{\sc Lightning}, our SED fitting code, is available to the astronomy community through the Astrophysics Source Code Library.

\subsection{Extinction Parameters}\label{sec:extinction}
As described above, we solve for the SFH step intensities by employing the matrix inversion procedure described in \S\,\ref{sec:chi2minimization}. To determine the extinction parameters and their uncertainties, we create a three-dimensional grid, with 81 steps in each dimension with
\begin{equation}
0.0 \le \tau_{{\rm DIFF},V} \le 4.0\ ,
\end{equation}
\begin{equation}
-2.3 \le \delta \le 0.4\ ,
\end{equation}
\begin{equation}
0.0 \le \tau_{{\rm BC},V} \le 4.0\ ,
\end{equation}
as in \cite{Leja2017}.
We adopt flat priors for all three parameters, and also impose the joint prior $\tau_{{\rm DIFF},V}/\tau_{{\rm BC},V}<2.0$ used by \cite{Leja2017}, to reproduce the previous observations suggesting that the total optical depth towards younger stars that ionize the nebular emission lines emitting gas is around twice that of the older stellar component \citep{Calzetti1994,Price2014}.

For each of the 531,441 (i.e., 81$^3$) discrete points of extinction parameter space grid, we employ the inversion procedure to determine the five $\psi_i$ coefficients of the SFH from the observed SED. For each of the 2043 pixels, we calculate the minimum $\chi^2$ for each point of this 3D grid of models.

With the current grid, we have performed both a Bayesian analysis in 3D extinction parameter space to derive marginalized likelihoods and posterior probability densities for each extinction parameter and a frequentist analysis repeating this minimization procedure for each of the 400 simulated SEDs, perturbed around the observed SED. No statistically significant difference was observed between the two approaches regarding the confidence intervals for each of the three extinction parameters.

Henceforth, we quote only frequentist uncertainties and confidence intervals, resulting from the fits of the 400 simulated maps we generated for each of the 2043 observed SEDs. Every quantity derived from the fits is measured 400 times, by selecting the model that maximizes the posterior probability distribution and not necessarily the model with lowest $\chi^2$ statistic. No significant difference is observed between the minimum $\chi^2$ and the $\chi^2$ for the model that maximizes the posterior probability density. For any derived quantity, we therefore quote the median value of the 400 simulations as the solution, with the lower and upper error bars corresponding to the differences to the 16th and 84th percentiles, respectively.

%

\section{SED Fitting Results}\label{sec:results}

\subsection{Derived Quantities}\label{sec:derived}
We have applied our SED fitting procedure to the 13 images of M51, which were registered to a common grid with 2043 square pixels of 10\arcsec (see \S\,\ref{sec:convolution} for details). Figure~\ref{fig:chisqr_map} displays the $\chi^2$ map for the model that maximizes the posterior probability for each pixel. Median $\chi^2$ considering all pixels is 6.8, which is very reasonable, considering we have 13 observables, 3 dust parameters, and 5 SFH steps (5 degrees of freedom). 

\begin{figure}
  \singlespace
  \centering
  \includegraphics[width=0.5\textwidth]{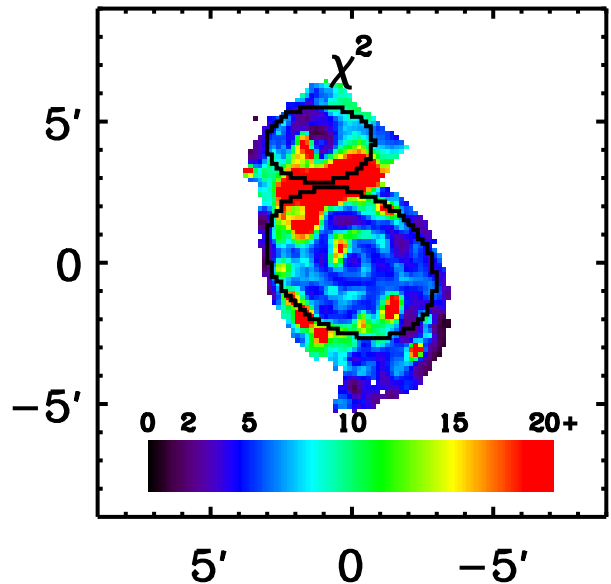}
  \vspace{-0.2in}
  \caption{Goodness-of-fit ($\chi^2$) map, corresponding to the models that maximize the posterior probability for each pixel. Ellipses are the same as those in Figure~\ref{fig:lir_map}}
  \label{fig:chisqr_map}
\end{figure}

Figure~\ref{fig:sed_medianchisqr} presents the SED fitting of the pixel with median $\chi^2$, with a highest posterior probability model having $\chi^2=6.8$. This exemplifies a typical fit. The observed SED is shown as the empty black diamonds, with the last point in the side panel corresponding to the TIR luminosity. All error bars on the black points are 1$\sigma$ uncertainties and include background subtraction and calibration uncertainties. The attenuated SED model is shown in red, and in blue we show the intrinsic, unattenuated SED model. The red point fitting the TIR luminosity is the difference in bolometric luminosity between the (high-resolution) intrinsic and attenuated model spectra, as in Equation~\ref{eq:ltir}.

\begin{figure}
  \singlespace
  \centering
  \hspace{-.3in}
  \includegraphics[width=0.50\textwidth]{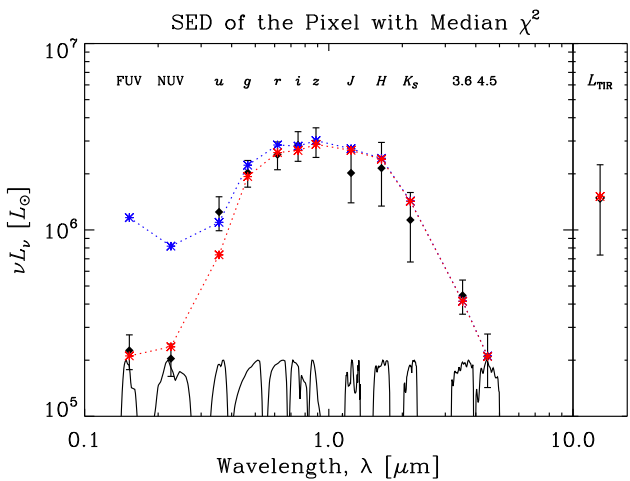}
  \caption{SED of the pixel with the median $\chi^2$ ($\chi^2_{\rm med}=6.8$) of all of our pixels. The observed luminosities are shown as the black empty diamonds. The observed TIR luminosity, combining the 24, 70, 160, and 250\,\um\ data is displayed in the side panel. The error bars correspond to the 1$\sigma$ uncertainties, which include background subtraction and calibration uncertainties. In red we display the model with maximum posterior probability and in blue we show its intrinsic SED (unattenuated). The filter response functions are shown as well. The attenuated power corresponds to the red model that fits the observed TIR luminosity.}
  \label{fig:sed_medianchisqr}
\end{figure}

The results for the four regions defined in Figure~\ref{fig:lir_map}, right are summarized in Table~\ref{tab:results_regions}. The table lists all five SFHs step intensities, the average SFR averaged over the last 100~Myr (${\rm SFR100}=0.1\psi_1+0.9\psi_2$), the total stellar mass $M_\star$ enclosed in each region, and the {\color{black}specific star formation rate (sSFR)}, ${\rm sSFR100}\equiv{\rm SFR100}/M_\star$.

\begin{deluxetable*}{ccccc}
\tabletypesize{\footnotesize}
\tablewidth{0in}
\tablecaption{SED Fittings Results for Regions}
\tablehead{ & M51a & M51b & outskirts & all }
\startdata
$\psi_1$ [\msunyr] & $ 2.4^{+1.0}_{-0.7}$ & $0.20^{+0.07}_{-0.06}$ & $0.45^{+0.13}_{-0.11}$ & $ 3.0^{+1.2}_{-0.8}$\vspace{0.06in}\\
$\psi_2$ [\msunyr] & $ 2.7^{+1.4}_{-1.0}$ & $0.08^{+0.03}_{-0.02}$ & $0.51^{+0.16}_{-0.13}$ & $ 3.3^{+1.7}_{-1.1}$\vspace{0.06in}\\
$\psi_3$ [\msunyr] & $10.3^{+2.8}_{-2.6}$ & $0.22^{+0.11}_{-0.11}$ & $1.77^{+0.38}_{-0.38}$ & $12.3^{+3.3}_{-2.9}$\vspace{0.06in}\\
$\psi_4$ [\msunyr] & $ 5.0^{+2.0}_{-1.8}$ & $2.46^{+0.31}_{-0.29}$ & $1.29^{+0.24}_{-0.25}$ & $ 8.8^{+2.0}_{-1.9}$\vspace{0.06in}\\
$\psi_5$ [\msunyr] & $ 4.0^{+2.0}_{-2.1}$ & $3.99^{+0.47}_{-0.44}$ & $0.24^{+0.15}_{-0.10}$ & $ 8.3^{+2.0}_{-2.1}$\vspace{0.06in}\\
SFR100 [\msunyr]   & $2.7^{+1.4}_{-0.8}$ & $0.09^{+0.03}_{-0.02}$ & $0.51^{+0.14}_{-0.12}$ & $ 3.3^{+1.5}_{-1.0}$\vspace{0.06in}\\
$M_\star$ [10$^{10}$\msun]& $3.4^{+0.4}_{-0.5}$ & $2.1^{+0.1}_{-0.1}$ & $0.49^{+0.03}_{-0.04}$ & $6.0^{+0.5}_{-0.6}$\vspace{0.06in}\\
log$_{10}$ sSFR100 [yr$^{-1}$]   & $-10.10^{+0.15}_{-0.15}$ & $-11.37^{+0.13}_{-0.12}$ & $ -9.99^{+ 0.09}_{- 0.10}$ & $-10.26^{+0.15}_{-0.14}$\vspace{0.06in}
\enddata
\label{tab:results_regions}
\end{deluxetable*}

Figure~\ref{fig:sfh} shows the SFHs of these four regions, using the same colors as the right panel of Figure~\ref{fig:lir_map}. A very interesting feature is evident from the outskirts region, implying that  the peak of star formation activity happened at the third SFH step, i.e., corresponding to stellar ages between 100\,Myr and 1\,Gyr. M51b has a predominantly old stellar population, with most of its mass coming from the oldest SFH step and considerably little SF activity over the last 100\,Myr. These results are consistent with the findings of \cite{MentuchCooper2012}, who modeled the SFH across the M51 system with two exponential decays and recovered an older stellar component with ages between 7 and 10~Gyr and a younger component with ages between 300 and 600~Myr, which they attributed to the interaction between the two galaxies.

\begin{figure}
  \singlespace
  \hspace{-.5in}
  \includegraphics[width=0.55\textwidth]{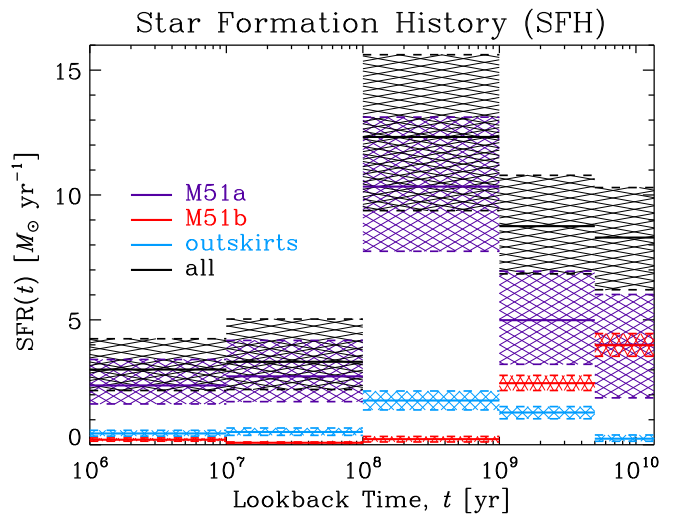}
  \vspace{-0.2in}
  \caption{Five-step SFHs of all pixels combined (black), the M51a ellipse (violet), the M51b ellipse (red), and pixels outside the two ellipses (cyan). The solid lines denote the median values from the simulations and dotted lines show the 16th and 84th percentiles, enclosing the hashed areas. The first SFH step is actually from 0 to 10~Myr, but it is displayed here from 1 to 10~Myr for simplicity. The pixels outside of the two ellipses (blue) correspond to the outskirts of the spiral galaxy M51a and the interaction region between M51a and M51b. Their combined SFH points to an enhancement of SFR not associated with the recent SFR (last 100\,Myr), corresponding to ages between 100 Myr and 5 Gyr (SFH steps 3 and 4). M51b seems to have the bulk of the oldest stellar population (fifth SFH step; ages between 5 and 13.6 Gyr) and a declining SFH. Also, the M51b ellipse does not contain considerable amounts of stars produced by the interaction. The spiral galaxy (M51a), however, contains mainly stars produced by the third and fourth SFH steps, likely strongly affected by the interaction. }
  \label{fig:sfh}
\end{figure}

\begin{figure*}
  \singlespace
  \includegraphics[width=\textwidth]{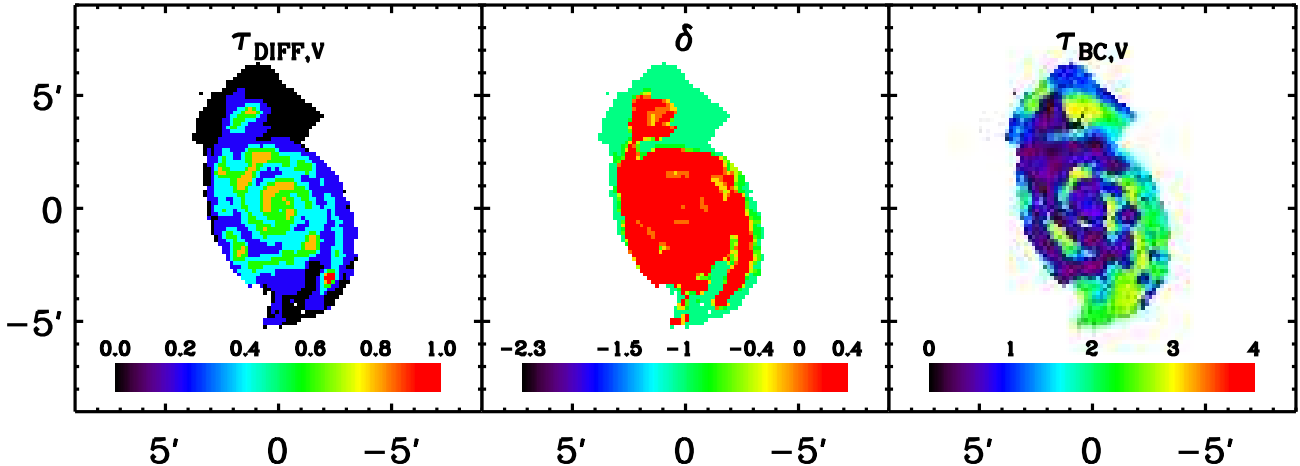}
  \vspace{-0.2in}
  \caption{Maps of the three extinction parameters. The left panel displays the optical depth at the $V$-band, $\tau_{{\rm DIFF},V}$, which is applied to SFH steps of all ages. In general attenuation is less than unity and it increases slightly where star formation is more intense in the last $\sim$100\,Myr. The middle panel displays the parameter $\delta$ that controls the slope of the extinction curve. The solutions favor curves around Calzetti extinction curve ($\delta=0)$ and strongly disfavor much steeper solutions ($\delta<-0.4$). Right panel shows the extra $V$-band optical depth for birth clouds, $\tau_{{\rm BC}, V}$, that applies only to the most recent SFH step, affecting stellar population younger than 10\,Myr. This component of the attenuation is also small, usually less than unity. It, however, rises to high values in the inter-arm regions, which does not mean the full attenuation in the inter-arm regions is high.}
  \label{fig:parameters_map}
\end{figure*}

\begin{figure}
  \singlespace
  \includegraphics[width=0.66\textwidth]{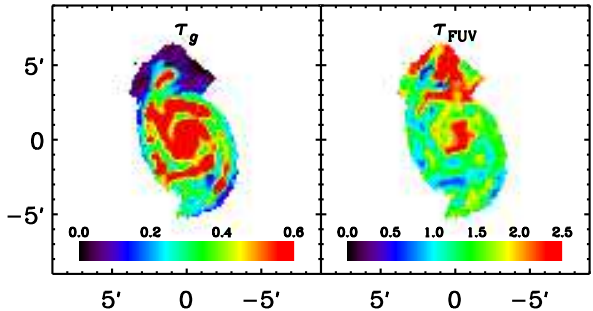}
  \vspace{-0.2in}
  \caption{Maps of the combined attenuation at the SDSS $g$ and \galex\ FUV bands. The recovered $g$-band attenuation traces arms and star-forming regions quite well. The FUV attenuation map is much messier. This can happen because of the higher uncertainties on the recovered FUV attenuation or because it indeed has a more complicated morphological structure.}
  \label{fig:tau_g_FUV}
\end{figure}

Figure~\ref{fig:parameters_map} displays the maps of the solutions for the three extinction parameters. The left panel shows the diffuse optical depth at the $V$-band, $\tau_{{\rm DIFF},V}$, which was applied to SFH steps of all ages. In general, attenuation is less than unity and it increases slightly where star formation is more intense in the last $\sim$100\,Myr. The middle panel displays the parameter $\delta$ that controls the slope of the extinction curve. The solutions favor curves around Calzetti extinction curve ($\delta=0)$ and strongly disfavor much steeper solutions ($\delta<-0.4$). The right panel shows the extra $V$-band optical depth for birth clouds, $\tau_{{\rm BC}, V}$, that applies only to the most recent SFH step, affecting the stellar population younger than 10\,Myr. This component of the attenuation is also small, usually less than unity. It, however, rises to high values in the inter-arm regions, which does not mean the full attenuation in the inter-arm regions is high. Figure~\ref{fig:tau_g_FUV} shows the final recovered attenuations at the SDSS $g$ and \galex\ FUV bands. The combined $g$-band attenuation traces the arms and star forming regions very well, but in the FUV the solution is much messier. This is likely due to larger propagated uncertainties in the FUV attenuation, but also because the FUV attenuation does have more complicated morphological structure.

Maps of the SFR densities $\Sigma(\psi_i)$ at different epochs are shown in Figure~\ref{fig:coeff_map}, in units of $\msunyr\,{\rm kpc}^{-2}$, as well as the average SFR density in the last 100~Myr, $\Sigma$(SFR100). The figure shows the stellar mass of M51b comes mainly from the 5th SFH step (ages from 5 to 13.6~Gyr), as expected from Figure~\ref{fig:sfh}, but M51a also has an old component mainly in its center. M51a experienced a lot of SF from between 100~Myr and 5~Gyr ago and the stars produced during this period are well mixed throughout the disk.

Interestingly, in the last 100~Myr, the timescale usually adopted to infer the current SFR from the UV emission, the SFR of M51 has been relatively weak. This implies a considerable fraction of the light in all photometric bands (including the UV) is arising from older stars, not associated with this ``recent" SFR.  

\begin{figure*}
  \singlespace
  \includegraphics[width=\textwidth]{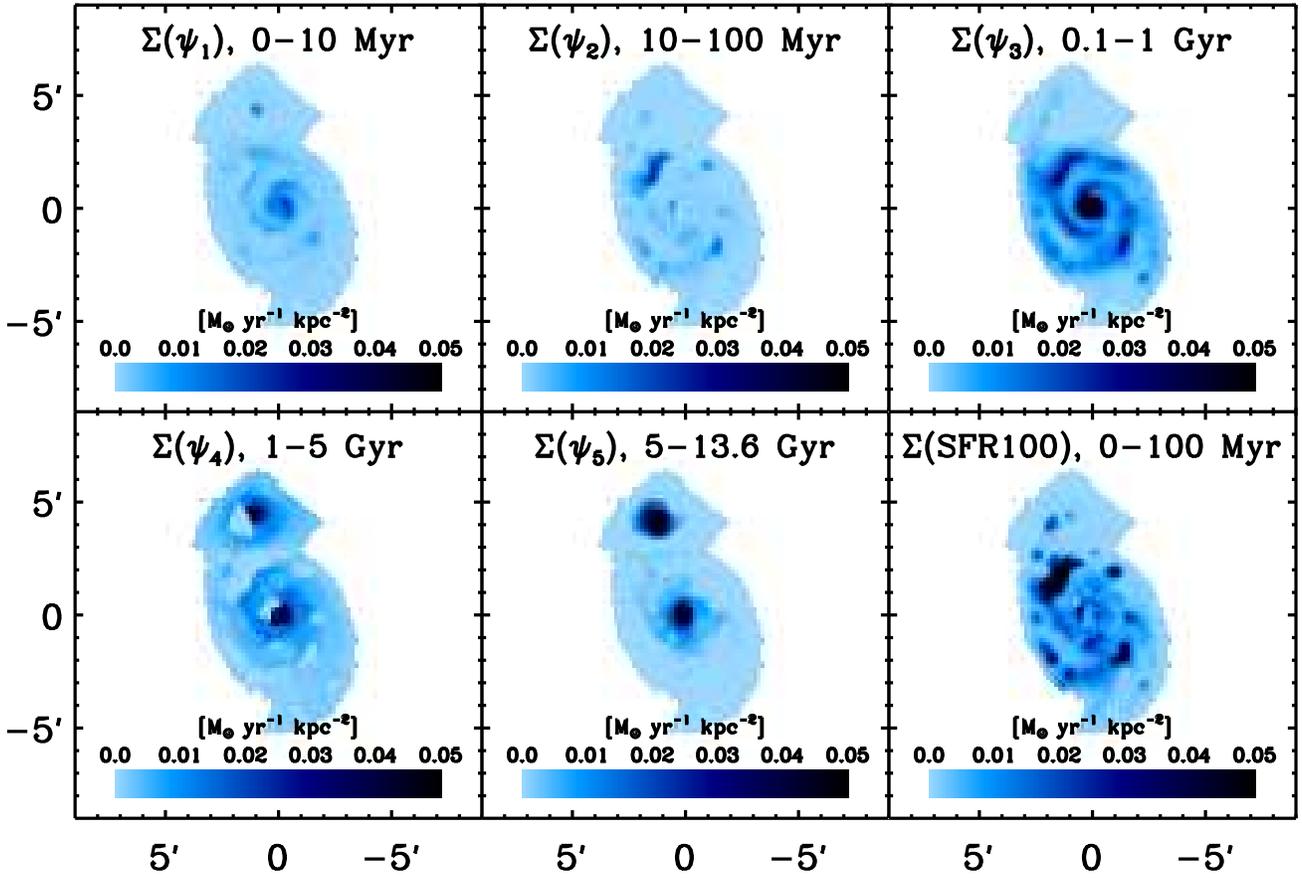}
  \vspace{-0.2in}
  \caption{Maps of the SFR surface density at different epochs. All maps in units of $\msunyr\,{\rm kpc}^{-2}$. Maps correspond to the highest likelihood models, but even for pixels where the best solution is zero, we can still provide upper limits and confidence intervals from our simulations.}
  \label{fig:coeff_map}
\end{figure*}

Our model naturally separates the contributions of each SFH step to any quantity derived by our model. We are able to, for example, quantify that the third SFH step (ages between 100~Myr and 1~Gyr) contributes {\color{black} $22.9^{+8.3}_{-7.5}$\%} of the total intrinsic FUV. The second step (10--100~Myr), which would naively be expected to have the most impact on the total FUV luminosity, contributes {\color{black}$24.1^{+13.6}_{-8.1}$\%}. 

We also estimate that $11.8^{+4.8}_{-3.3}$\% of the stellar mass of the whole map arises from the third SFH step, very consistent with the burst stellar mass fraction of between 5--15\% derived by \cite{MentuchCooper2012}. This shows the effect of the interaction on the system as a whole. Our model indicates $36.3^{+ 9.6}_{-14.3}$\% of the current stellar mass coming from the fourth, and $51.5^{+13.9}_{-10.2}$\% from the fifth, oldest step. For completeness, only $0.049^{+0.019}_{-0.017}$\% and $0.34^{+0.28}_{-0.10}$\% of the total current stellar mass were produced by the first and second SFH steps, respectively.

\subsection{{\sc Lightning} Parameter Recovery}\label{sec:recovery}
We performed simulations of parameter recovery for a variety of theoretical SFHs, including scenarios where SFR increases or decreases with lookback time, as well as step functions where only one SFH step is non-zero, and cases with a sharp drop. As an example of parameter recovery and of the degeneracies involved, we use the best-fit SFH and extinction parameters derived for the large region M51a as simulation inputs. This large region (with 862 pixels) is detected with high confidence over the background noise in all images and its SED uncertainties arise mainly from calibration uncertainties on the photometry (see Table~\ref{tab:dataset}).

The observed SED was fitted once, then we simulated 400 SEDs around the best-fit model SED. The simulated SEDs were generated as described in \S~\ref{sec:chi2minimization}. Figure~\ref{fig:M51a_plottile} is one of the diagnostic outputs of {\sc Lightning}. It shows the results of this simulation. First, histograms of all 8 parameters were plotted. Then the maximum value of each $x$-axis was used to discretize the full 8D parameter space. For each of the five SFH intensities, we created 1D grids with with 15 values, linearly increasing from 0 to the respective maximum value. The 3D extinction grid has respectively 21, 14, and 21 values for $\tau_{{\rm DIFF},V}$, $\delta$, and $\tau_{{\rm BC},V}$. The whole 8D grid has therefore $\approx$4.7 billion models ($15^5\times21\times14\times21$).

In Figure~\ref{fig:M51a_plottile} we also present histograms of the 400 results for the best-fit $\chi^2$ values, SFR100, stellar mass, bolometric luminosity, and sSFR100 are also shown, with the median results as blue dashed vertical lines and the best fits to the data as red vertical lines. Figure~\ref{fig:M51a_plottile}  also includes a set of scatter plots with the results of best-fits to the 400 simulations of the SED (clouds of points), the median values recovered from the simulation (empty diamonds), the best-fit to the data (solid, red stars), and the various Bayesian marginalized likelihoods for eight parameters (five SFH step intensities and three extinction parameters, shown as the red contours and curves). The three contour levels enclose 68\%, 95\%, and 99\% of the 2D marginalized likelihoods.

Various degeneracies are at play. Figure~\ref{fig:M51a_plottile} shows how well each parameter and derived quantity is recovered both from the Bayesian and from the frequentist analysis. The results from both methods agree, with the best-fit to the data (shown as red stars or red vertical lines) matching, within uncertainties, the median values from the 400 simulations (shown as blue diamonds or dashed vertical lines). Many quantities, however, are derived effectively as upper limits, in agreement with the findings of \cite{Leja2017}, who used a Markov Chain Monte Carlo (MCMC) method. For this particular SED, $\psi_3$ and $\tau_{{\rm DIFF},V}$ are well recovered and both distributions agree for these parameters. The final likelihood is almost independent of $\tau_{{\rm BC},V}$, with lower values around 0.5 being weakly favored, and the favored extinction curve is flatter than the Calzetti curve ($\delta\sim0.2$). All other parameters ($\psi_1$, $\psi_2$, $\psi_4$, and $\psi_5$) receive proper upper limits and the best-fit to the data agree with the median from the simulations, within uncertainties. The $\chi^2$ distribution shows the recovered goodness-of-fit for the data is reasonable, with the model not overfitting ($\chi^2$ too low) nor being inadequate ($\chi^2$ too high).

We also show the histograms for the derived SFR100, stellar masses, bolometric luminosities, and sSFR100, as derived from the 400 simulations. We do not show these marginalized Bayesian distributions, as computing them is too computationally intensive. In fact, one of the main advantages of {\sc Lightning} is generating a vector of derived quantities with the same size of the number of simulations, not having a burning phase as it would happen to MCMC methods.

In this work, {\sc Lightning} was extensively tested with real data, considering the large variety of SEDs fitted here (2043 pixels and 4 large regions). In the next section, we compare the results of {\sc Lightning} with a widely tested SED fitting code.


\begin{figure*}
  \singlespace
  \includegraphics[width=\textwidth]{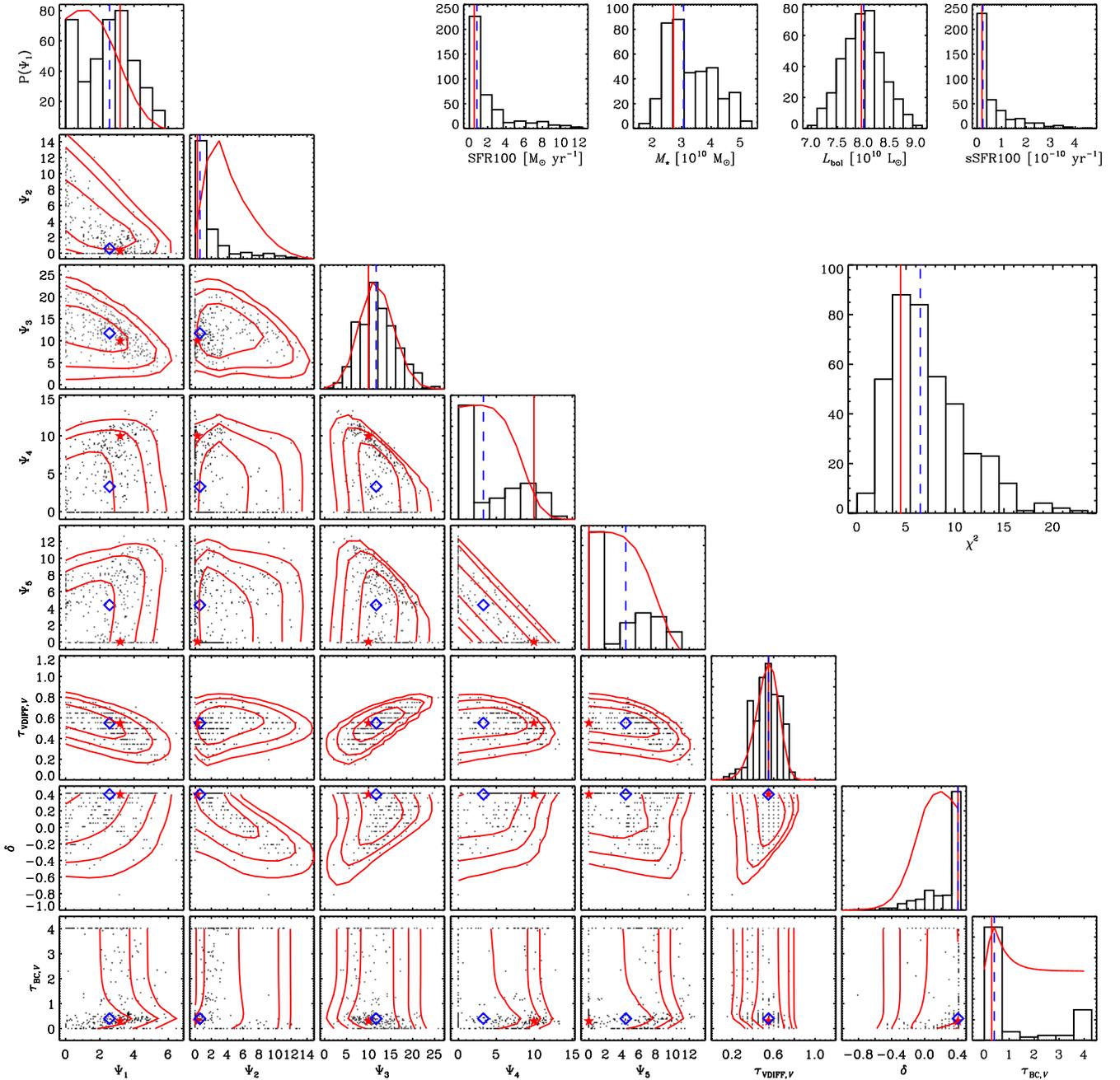}
  \vspace{-0.2in}
  \caption{Diagnostic plots for the fitting of the SED of the large region M51a. Scatter plots of the values recovered from the 400 simulated SEDs are shown for the two-by-two combinations os parameters and histograms are shown for all eight parameters, SFR, stellar mass, bolometric luminosity, sSFR, and best-fit chi-squares. The best-fit values to the observed data in the histograms as the red, solid, vertical lines (or red stars) and the median values from the 400 simulations are shown as the blue, dashed, vertical lines (or blue diamonds). On top of the histograms for all eight parameters we display Bayesian 1D marginalized likelihoods. All two-by-two combinations of 2D scatter plots are shown with the 2D marginalized likelihoods overlaid as the red contours (enclosing 68\%, 95\%, and 99\% of the likelihood).}
  \label{fig:M51a_plottile}
\end{figure*}

\subsection{Comparison to CIGALE}\label{sec:comparisons}

We have compared our results with the well-tested, widely used Python implementation \citep{Roehlly2014,Boquien2014} of the CIGALE SED modeling code \citep{Burgarella2005,Noll2009}, PCIGALE. We ran the latest version (0.11.0) of the code on the SEDs of all 2043 pixels and of the four separate large regions.

Initially, we ran CIGALE with the same grid as \cite{Boquien2014}. \cite{BC03} stellar population models were used, assuming a \cite{Chabrier2003} and a solar metallicity ($Z=0.02$). The SFH consists of two exponentially decaying populations, an old stellar population with formation age of 13~Gyr and a young stellar population with eight possible formation ages (5, 10, 25, 50, 100, 200, 350, or 500~Myr). For a complete list of the parameter values used in CIGALE, see Table~2 of \cite{Boquien2016}. The FIR SEDs are fitted with \cite{Dale2014} templates. Excellent agreement is found between the TIR luminosities derived from \cite{Galametz2013} calibration based on \cite{DL2007} models (Equation~\ref{eq:lir}) and the ones derived with CIGALE, with the CIGALE values being only 5\% larger (less than the calibration uncertainties) on average, considering all 2043 pixels.

With this initial model, however, we observe a significant discrepancy between the stellar masses recovered from CIGALE and {\sc Lightning}, with the best-fit masses from CIGALE being 47\% higher on average, considering all 2043 pixels. We suspected this disagreement was most likely due to the fixed 13~Gyr age for the older stellar population used by \cite{Boquien2016}, which has considerably lower mass-to-light ratio. We added more freedom for the possible ages of the older stellar population. CIGALE accepts at most three ages for the old stellar population. We then chose 5, 8, and 13~Gyr as the new age grid for the old stellar population. We also added 1 and 2~Gyr as possible ages for the young stellar population. After this simple modification, reasonable agreement is reached between the stellar masses derived from {\sc Lightning} and CIGALE, with the pixel-by-pixel stellar masses from CIGALE being only 2.2\% higher on average.

To compare the SFHs derived from CIGALE with those derived from {\sc Lightning}, we averaged individually the high-resolution SFHs of all 2043 pixels over the five time bins considered here. Figure~\ref{fig:SFH_comparison} shows the results of this comparison adding the SFHs over the four large regions considered in this work. SFHs for M51a, M51b, outskirts, and all pixels combined are shown, respectively, in violet, red, cyan, and black. High-resolution models from CIGALE are shown as the solid lines and the thicker solid lines show the high-resolution CIGALE models averaged over our five time bins. Note many exponential decays are seen in a single region's SFH model. This happens because there are hundreds of pixels in each region, with the SFH of each pixel containing two exponential decays of different decay timescales. The number of pixels in each region is actually listed in Table~\ref{tab:regions}. 1--$\sigma$ confidence intervals (between the 16th and 84th percentiles) from {\sc Lightning} are shown as the hashed areas and the median values are shown as the dotted lines.

\begin{figure*}
  \singlespace
  \includegraphics[width=\textwidth]{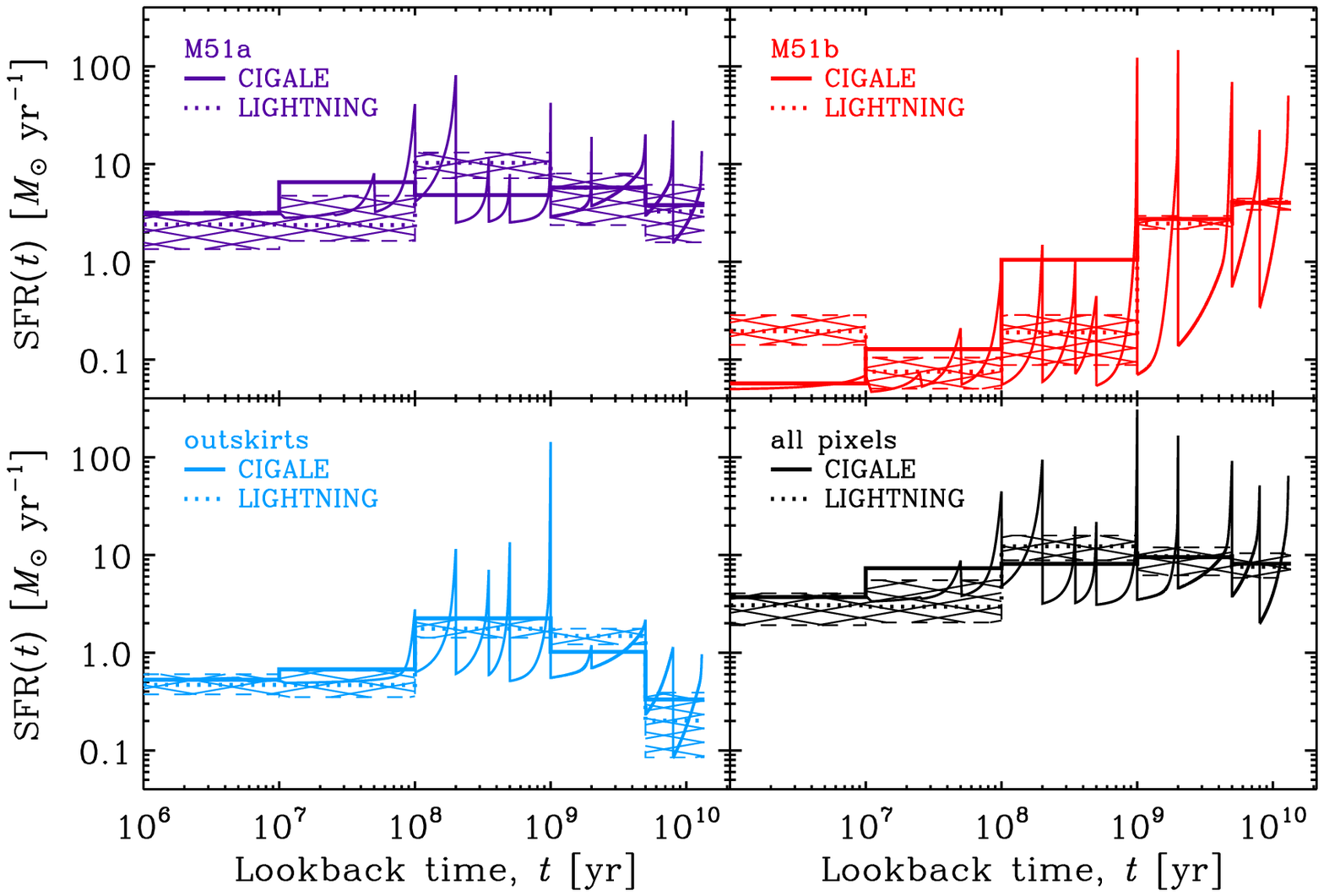}
  \vspace{-0.2in}
  \caption{SFHs from {\sc Lightning} compared to the ones derived from CIGALE. SFHs were derived for all 2043 pixels separately and then added for the four spatial regions considered here. SFHs for M51a, M51b, outskirts, and all pixels are shown in violet, red, cyan, and black, respectively. Hashed areas show 1--$\sigma$ confidence intervals (between 16th and 84th percentiles) as derived with {\sc Lightning}. Dotted lines show median {\sc Lightning} values. Thick solid lines show best-fit CIGALE SFHs averaged over the time bins used here. Solid curves show best-fit high-resolution CIGALE models. Uncertainties from CIGALE are not shown.}
  \label{fig:SFH_comparison}
\end{figure*}

Note these are the best-fit SFHs from CIGALE, and no uncertainties were derived on these high-resolution SFHs, for the simple reason that effort would be too computationally intensive. Qualitatively, however, similar trends are recovered from the two models. Considering CIGALE's uncertainties are of the same magnitude as the ones derived from {\sc Lightning}, there is virtually no significant discrepancy. For example, one apparently large disagreement is the 0--10~Myr SFR in M51b, which is lower by a factor of 3 or 4 in Figure~\ref{fig:SFH_comparison}. However, with the simpler, initial CIGALE grid the Bayesian SFR10 (averaged over the last 10~Myr) is 0.114$\pm$0.018\,\msunyr, while the best-fit SFR is 0.16\,\msunyr, and therefore in excellent agreement with our estimate. We have good reason to believe that a more detailed CIGALE model accepting more mass ratios between the younger and the older would bring CIGALE's answer closer to {\sc Lightning}'s. This increase in complexity to fix such a small detail is beyond the scope of this work. Since the SFH steps were chosen here to have similar bolometric luminosity per unit SFR, the large stellar mass in the older SFH bin for M51b outshines the younger SFH bin by at least a factor of 10. Of course, in the FUV the youngest first SFH step still outshines the oldest, fifth step by a factor of 40 and the SFR can be recovered.

All in all, remarkable agreement is found between the results of CIGALE and {\sc Lightning}, especially considering the two codes use different evolutionary population synthesis models, (slightly) different IMFs, dust extinction treatment, and fitting techniques. Regarding running time, however, for an extinction grid with as many models as CIGALE, {\sc Lightning} performs $\sim$2000 times faster. We point out CIGALE models the IR SED and {\sc Lightning} is using the TIR luminosities from \cite{Galametz2013}, which is responsible for part of the speedup. {\sc Lightning} derives fast and reliable SFHs, which meets our current goal, but we plan to add IR SED fitting in a later version of {\sc Lightning}.

\section{Discussion}\label{sec:discussion}

\subsection{Hybrid Star Formation Rate Across M51}\label{sec:hybridSFR}
Here we set out to use our SED fitting results to derive hybrid SFR laws combining UV and IR indicators. As an example, we show the \galex\ FUV + \spitzer\ 24\,\um, but the reasoning used here can be applied to all other hybrid SFR laws. 

The UV has the convenient property of tracing almost exclusively young stellar populations. If a galaxy, or a region of a galaxy, forms stars at a constant rate, its UV luminosity reaches a plateau and that plateau is commonly used as a SFR calibration. The FUV reaches a steady state (i.e., constant $L_{\rm UV}/{\rm SFR}$) on timescales of 100~Myr. The same can be said for tracers of ionizing photon rate, like H$\alpha$, H$\beta$, and Paschen~$\alpha$ line emission intensities, hydrogen recombination continuum intensities, or free-free intensities, but that plateau timescale is much shorter, usually less than 10~Myr. Qualitatively, these timescales correspond to lifetimes of the least massive stars that produce significant UV or ionizing photons. \cite{Eufrasio2015} tabulated these timescales for different bands, for various IMFs, and metallicities. One main issue that arises when modeling SFHs is that an inappropriate choice of models might not be able to properly recover independently the UV and H$\alpha$ fluxes. Interestingly, \cite{Leja2017} shows that one is able to recover the proper H$\alpha$ fluxes by solely modeling broadband SEDs with step SFHs, very similar to the ones used here. 

Based on the above, the UV emission is a strong tracer of the recent SFR, and the conversion from a given UV band intensity to SFR can easily be derived from stellar population synthesis. These conversions depend on IMF and metallicity, as well as the timescale considered for the SFR scenario. From the population synthesis code {\sc P\'egase}, assuming a Kroupa IMF, a stellar population with solar metallicity ($Z=0.02$) that formed stars uniformly during the last 100\,Myr (and nothing before) will have a SFR given by
\begin{equation}\label{eq:sfr_fuv_100Myr}
  \left(\frac{\rm SFR100}{\msunyr}\right) = 1.6\times10^{-10}\left(\frac{L{\rm _{FUV}^{intr}}}{\lsun}\right)
\end{equation}
and
\begin{equation}\label{eq:sfr_nuv_100Myr}
  \left(\frac{\rm SFR100}{\msunyr}\right) = 2.7\times10^{-10}\left(\frac{L^{\rm intr}_{\rm NUV}}{\lsun}\right)
  {\rm.}
\end{equation}

More generally, one can write
\begin{equation}\label{eq:sfr_x}
  \left(\frac{\rm SFR100}{\msunyr}\right) \equiv k_{X}\left(\frac{L{\rm _{X}^{intr}}}{\lsun}\right),
\end{equation}
where $k_{X}$ is the conversion from the intrinsic luminosity of band $X$ to the average SFR over the last 100\,Myr \citep{Kennicutt1998,Kennicutt2012,Murphy2011}.

Of course, these conversions rely on the assumption of constant SFR over the last 100~Myr and nothing before it. Looking at Figure~\ref{fig:sfh}, however, one might expect that this is not the case for any of our regions. Actually for the FUV calibration conversion $k_{\rm FUV}$ (Equation~\ref{eq:sfr_fuv_100Myr}), instead of the factor 1.6 for the simple, idealized case, the SFHs derived here for M51a, M51b, outskirts, and all regions lead to conversion factors of 
{\color{black}
$1.29^{+0.89}_{-0.39}$, 
$0.81^{+0.39}_{-0.17}$, 
$1.33^{+0.66}_{-0.30}$, and 
$1.28^{+0.83}_{-0.37}$
}, respectively. For the NUV band, instead of 2.7, the conversion factors are 
{\color{black}
$1.72^{+1.22}_{-0.52}$,
$1.06^{+0.51}_{-0.21}$,
$1.77^{+0.86}_{-0.39}$, and 
$1.71^{+1.17}_{-0.50}$
}. Thus the simplified assumption yields a $k_X$ higher by 40\%--80\% for M51, which alone would overestimate the SFR. Note that the conversion being different is not a surprising result of our fits, but is expected from a theoretical standpoint, if the SFH is anything other than a constant SFR for the last 100~Myr preceded by no SFH. It just happens that the SFHs across M51 are quite different from the simplifying assumption. We therefore emphasize the importance of properly modeling the SFH and attenuation parameters in order to derive reliable SFR calibrations.

One difficulty of using the UV luminosities as SFR indicators is that the UV fluxes are highly reprocessed by dust and gas, mainly due to dust attenuation and gas absorption and scattering. Some modeling is necessary to obtain intrinsic UV luminosities. SED fitting, as employed here, with as many observables as possible, is most likely one of most accurate ways of determining dust attenuation and emission.

Several authors, including \cite{Calzetti2007}, \cite{Leroy2008}, \cite{Zhu2008}, \cite{Kennicutt2009}, \cite{Hao2011}, and \cite{Lee2013}, have developed hybrid star-formation tracers combining UV or H$\alpha$ with IR emission.

Standard prescriptions to correct the UV luminosities for dust extinction are widely available, empirically calibrated for large samples of galaxies.
For instance, some commonly used include:
\begin{align}
  L_{\rm FUV}^{\rm corr}
  & = L{\rm _{FUV}^{obs}} + 3.89\times L{\rm _{24\mu m}^{obs}},         \label{eq:FUV_corr_24}\\
  & = L{\rm _{FUV}^{obs}} + 0.46\times L{\rm _{TIR}^{obs}},              \label{eq:FUV_corr_IR}
\end{align}
and 
\begin{align}
  L_{\rm NUV}^{\rm corr}
  & = L{\rm _{NUV}^{obs}} + 2.26\times L{\rm _{\rm 24\mu m}^{obs}},    \label{eq:NUV_corr_24}\\
  & = L{\rm _{NUV}^{obs}} + 0.27\times L{\rm _{TIR}^{obs}},             \label{eq:NUV_corr_IR}
\end{align}
as derived by \cite{Hao2011}.

After applying these prescriptions, the so-called ``corrected" luminosities on the left-hand sides of Equations~\ref{eq:FUV_corr_24}--\ref{eq:NUV_corr_IR} are treated as the intrinsic. These luminosities can then be converted into SFRs following the appropriate relations. Generalizing the equations above, we can define
\begin{equation}\label{eq:acorr_general}
  L_X^{\rm corr} \equiv L_X^{\rm obs} + a_{\rm corr}(X,Y)\times L_Y^{\rm obs},
\end{equation}
where $a_{\rm corr}(X,Y)$ is the correction factor multiplying observable luminosity $L_Y$ in order to account for the extinction in observable luminosity $L_X$. In fact, $a_{\rm corr}(X,Y)$ is the ratio of the attenuated light in the $X$ band over the intensity of the $Y$ band, i.e., the efficiency of attenuation in $X$ compared to emission of $Y$. 
\begin{align}
  \left(\frac{\rm SFR}{\msunyr}\right)
   & = k_{X}\left(\frac{L{\rm _{X}^{intr}}}{\lsun}\right) \\
   & = k_{X}\left[\left(\frac{L_X^{\rm obs}}{\lsun}\right) + 
            a_{\rm corr}(X,Y)\left(\frac{L_Y^{\rm obs}}{\lsun}\right)\right],
\end{align}
\begin{figure*}
  \singlespace
  \vspace{-.3in}
  \includegraphics[width=0.95\textwidth]{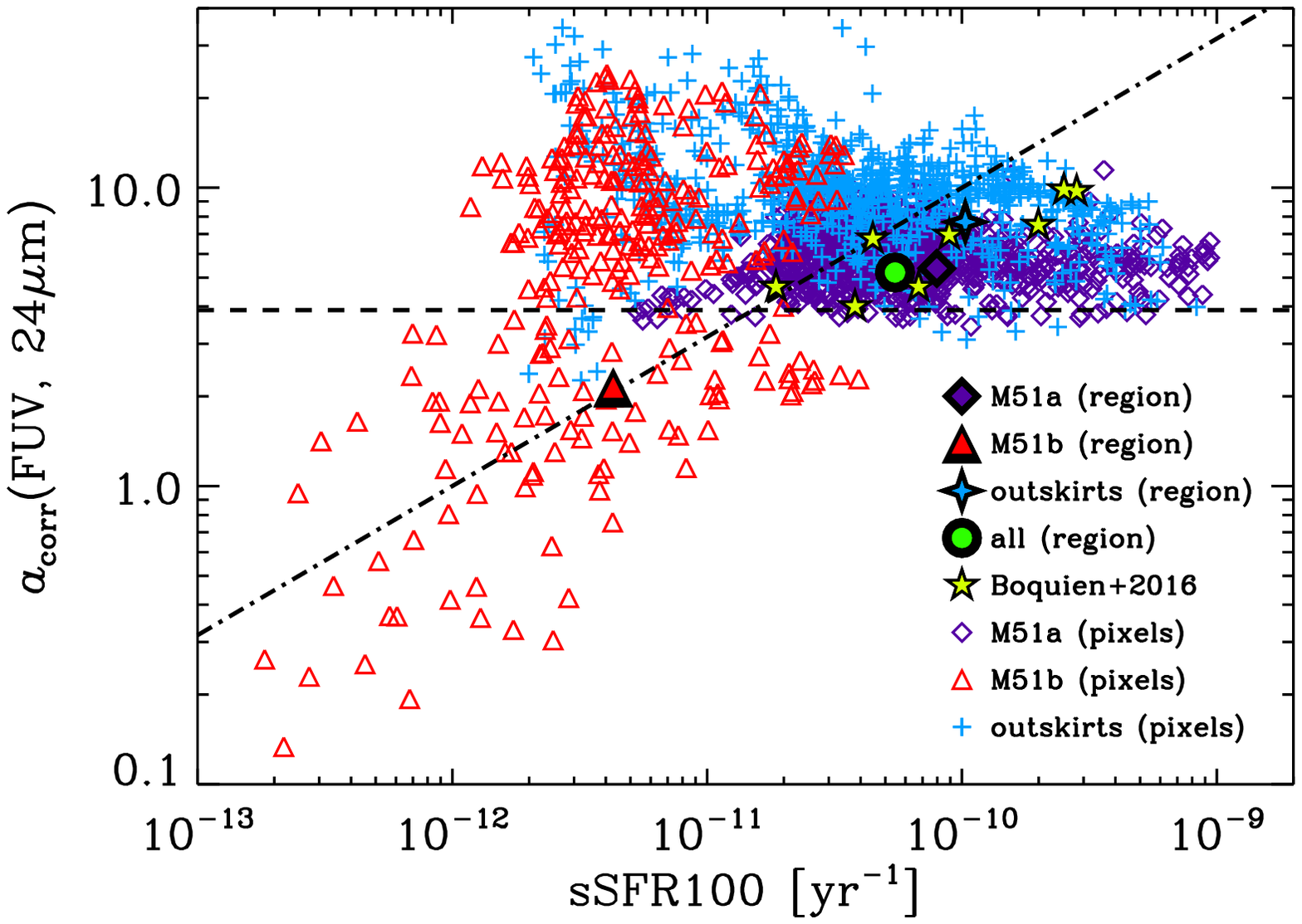}
  \vspace{-0.2in}
  \caption{The correction factor $a_{\rm corr}({\rm FUV},24\,\um)$ versus sSFR100 for all pixels of our map. Violet diamonds correspond to pixels inside M51a elipse, red triangles are inside M51b elipse, and cyan crosses are pixels in the ``outskirts" region. Global values for the three regions are displayed in larger, filled symbols. Dashed horizontal line shows \cite{Hao2011} correction factor and dotted line represents \cite{Eufrasio2014} relation. The eight galaxies from \cite{Boquien2016} are shown as filled stars. The fact that this correction factor changes with sSFR100 indicates the old stellar population, not associated with the recent SFR100, significantly contributes to the IR emission.}
  \label{fig:acorr_FUV_024_vs_sSFR100}
\end{figure*}

In \cite{Eufrasio2014}, we tested Equation~\ref{eq:acorr_general} combining the observed \galex\ FUV and \wise\ 22~\um\ emission (which is almost numerically identical to Equation~\ref{eq:FUV_corr_24}). The correction factor \acorr(FUV,~22\,$\mu$m) was observed to vary significantly across the giant interacting Condor galaxy, NGC~6872, for regions 10~kpc in diameter. \acorr(FUV,~22\,$\mu$m) were also observed to increase with increasing sSFR, meaning younger, bluer regions were observed to have higher attenuation at FUV wavelengths compared to emission at 22\,\um. However, FIR maps were not available for the Condor galaxy and energy balance was not imposed on those SED fittings. This made it more difficult to properly model the intrinsic FUV.

\cite{Boquien2016} modeled the spatially resolved SFHs of eight nearby face-on spiral galaxies with CIGALE and derived similar relations, where a variety of these correction factors, $a_{\rm corr}(X,Y)$, are clearly increasing with increasing sSFR, especially $a_{\rm corr}({\rm FUV},{\rm TIR})$. However, they seemed to also be correlated with the attenuations, stellar masses, and SFRs, albeit more weakly, and it was extremely difficult to determine if the trend was driven by the underlying stellar population or by geometry and differential attenuation effects.

These hybrid SFR tracers are widely used, since they are often the only way to determine SFRs for galaxies in the distant universe. Hence it is important to properly determine how the correction factor depends on attenuation and on the underlying stellar population properties.

Reliably measuring these correction factors requires good estimates of the intrinsic UV luminosity, which makes our model very suitable to tackle this problem. Taking advantage of SFRs derived from our previously described SED modeling, we produced maps of various correction factors, $a_{\rm corr}(X,Y)$, combining the \galex\ bands with the four FIR bands available for M51 (24, 70, 160, 250\,\um) and the TIR map. Figure~\ref{fig:acorr_FUV_024_vs_sSFR100} displays $a_{\rm corr}({\rm FUV},24\,\um)$ versus ${\rm sSFR100} = {\rm SFR100}/M_\star = \Sigma({\rm SFR100})/\Sigma(M_\star)$ for all our pixels. Empty violet diamonds correspond to pixels inside the M51a ellipse, empty red triangles are inside the M51b ellipse, and cyan crosses are pixels in the ``outskirts" region. The dashed and dotted lines correspond, respectively, to the \cite{Hao2011} and \cite{Eufrasio2014} relations. Treating all pixels in a region as a single location, we can also plot global quantities for our regions. The larger, filled symbols in the figure were added for the global quantity derived for each region, where the filled diamond, triangle, cross, and circle correspond to M51a, M51b, outskirts, and all regions combined, respectively. The eight galaxies from \cite{Boquien2016} are shown as filled yellow stars.

The problem of estimating recent SF activity from a combination of a UV and a FIR band is very complicated, with $a_{\rm corr}({\rm FUV},24\,\um)$ increasing on average with increasing FUV attenuation or SFR100, and increasing with decreasing stellar mass density. The fact that $a_{\rm corr}({\rm FUV},24\,\um)$ changes at all suggests that stars older than 100~Myr (not associated with the recent SFR traced by the UV) are capable of heating dust and significantly contribute to all FIR bands and to the TIR luminosity. 

In fact, \cite{Kennicutt2009} arrived to the conclusion that a considerable fraction of the TIR radiation in normal star-forming galaxies is heated by stars older than 100\,Myr. Their comparisons with evolutionary synthesis models suggested that up to 50\% of the TIR emission could be from dust heated from this evolved stellar population. They issued a stern warning that hybrid SFR indicators calibrated by them, \cite{Calzetti2007}, and \cite{Zhu2008} should not be applied to map the spatially resolved SFR in galaxies without the risk of introducing significant and possibly large systematic errors in the resulting SFR maps. These systematic uncertainties are exactly what we characterize across M51 in Figure~\ref{fig:acorr_FUV_024_vs_sSFR100}. 

Our model is perfectly suited to quantitatively check the hypothesis that old stars contribute to the FIR emission, since by construction we have five steps of the SFH probing different epochs. In the next section, we decompose the FIR emission into contributions from the five SFH steps.

\subsection{Decomposition of the IR Emission Into Contributions From The SFH}
\label{sec:ir_decomposition}

A significant body of work is available in the literature on the contribution of old stars to the IR emission of nearby galaxies. Various \herschel\ studies show warm dust to be associated with the average SFR of the last 100\,Myr and colder dust to be associated to older stars \citep{Popescu2000, Bendo2010, Bendo2012, Bendo2015, Boquien2011, DeLooze2012a, DeLooze2012b, Lu2014, Natale2015}. From hydrodynamical simulations, even in actively star-forming galaxies, up to a third of the TIR luminosity is associated with stars older than 100\,Myr \citep{Boquien2014,Boquien2016}. There is also strong evidence that the FUV$-$NUV color is strongly affected by the presence of old stars \cite{Boselli2005, Cortese2008}.

Disentangling IR emission from all ages is crucial to determine what fraction is associated with the recent SFR. \cite{Crocker2013} investigated the SFH effect on the TIR fraction heated by stars older than 100\,Myr. They found the result strongly depended on the SFH and extinction prescription used. For an exponentially decaying SFH of decay timescale of 5\,Gyr about half the TIR would be heated by old stars, while for a constant SFH, that number would be about a third.

From our model, we can directly determine the fraction of FIR emission coming from each SFH step of each pixel, since the TIR emission was one of the points of our generalized SEDs and it was fit using the linear combination of the five SFH components.
However, to better understand the contribution of the old stellar population to the IR emission, we can directly decompose each of the observed FIR images, as well as the TIR map, into contributions from our five SFH steps. 

We implemented our procedure using each of the large regions (i.e., M51a, M51b, outskirt, and whole map). We performed a $\chi^2$ minimization using the same inversion method used in our SED fittings (see \S\ref{sec:chi2minimization}). This time, however, we assumed that any of the observed IR band luminosities $L{\rm _{Y}^{obs}}$ were generated by a linear combination of the five derived SFH maps.

For instance, the model intensity of the $j$-th pixel in any region $S$ of the 24\,\um\ map is given by
\begin{equation}\label{eq:lir_decomp}
  L_{24,j}^{\rm mod} = \sum_{i=1}^5 k_i M_{\star,i,j},
\end{equation}
where this time $M_{\star,i,j}$ is fixed and the five $k_i$ constants are unknown. $M_{\star,i,j}$ is the current stellar mass of the $i$-th SFH step of the $j$-th pixel, which is proportional to $\psi_{i,j}$ (proportionality constants listed in Table~\ref{tab:sfh_facts}, for unit SFRs). The $k_i$ constants are therefore the mass-to-light ratios.

We then sum Equation~\ref{eq:lir_decomp} over a given region $S$ and rewrite it as
\begin{align}
  1 &= \sum_{i=1}^5 k_i \frac{\sum_{j \in S}M_{\star,i,j}}
                             {\sum_{j\prime \in S}L_{24,j\prime}^{\rm mod}}\nonumber\\
    &= \sum_{i=1}^5 k_i \frac{M_{\star,i,S}}{L_{24,S}^{\rm mod}}\nonumber\\
    &= \sum_{i=1}^5 f_{i,24,S},
\end{align}
where $f_{i,24,S} \equiv k_i M_{\star,i,S}/L_{24,S}^{\rm mod}$ is the fraction of the 24\,\um\ luminosity from region $S$ arising from the $i$-th SFH step. $M_{\star,i,S}=\sum_{j\in S}M_{\star,i,j}$ and $L_{24,S}^{\rm mod}=\sum_{j\in S}L_{24,j}^{\rm mod}$ are, respectively, the total stellar mass and the total 24\,\um\ luminosity from region $S$, associated with the $i$-th SFH step.

The goodness-of-fit $\chi^2$ can be written as
\begin{align}\label{eq:lir_decomp_chisqr}
  \chi^2 &= \sum_{j \in S} \frac{\left(L_{24,j}^{\rm mod} - L_{24,j}^{\rm obs}\right)^2}
                                {\sigma_j^2} \nonumber\\
         &= \sum_{j \in S} \frac{1}{\sigma_j^2}
                           \left(\sum_{i=1}^5f_{i,24,S}\frac{M_{\star,i,j}L_{24,S}}{M_{\star,i,S}} - L_{24,j}^{\rm obs}\right)^2.
\end{align}
Note the procedure can be applied to any IR broadband image of the galaxy replacing the 24\,\um. Also, Equations~\ref{eq:lir_decomp} and \ref{eq:lir_decomp_chisqr} are very similar to Equations~\ref{eq:sed_model} and \ref{eq:sed_model_chisqr}, but here we have as many observables as pixels in region $S$ and still five unknown fractions. The number of pixels in each region is listed in Table~\ref{tab:regions}. We therefore apply our previously discussed fitting procedure in order to solve for the five fractions, for each region and each IR band.

Table~\ref{tab:ir_fractions} lists the derived fractions $f_{i,\lambda,S}$ for all IR bands and regions. The largest contributor is in bold to guide the eye. The result is very surprising. For M51a the majority of the IR emission is coming from the stellar populations associated with the third SFH step, with ages between 100~Myr and 1~Gyr. For M51b, most of IR emission is mainly coming from the oldest stellar populations, with older than 5~Gyr. The outskirts region has most of the IR coming from the populations younger than 10~Myr, but comparable fraction arises from populations of ages between 100~Myr and 1~Gyr.

Thanks to the diversity of physical properties around M51, many of them influenced by the interaction between M51a and M51b, we observe different mixes of young and old stars, subjected to different attenuations. This diversity of IR fractions presented in Table~\ref{tab:ir_fractions} shows that even though the young stars do heat dust much more efficiently per stellar mass, a much larger population of old stars may compete (as in M51a) and even dominate the IR emission (as in M51b).

\begin{deluxetable*}{lccccccccccc}
\tabletypesize{\small}
\tabletypesize{\footnotesize}
\tablewidth{0in}
\tablecaption{FIR Luminosity Fractions,$f_i$, and Light-to-Mass Ratios, $k_i$, for each SFH Step and Region}
\tablehead{ Region & $\lambda_0$ & $f_1$ & $f_2$ & $f_3$ & $f_4$ & $f_5$&   $k_1/10$    & $k_2$         & $10\times k_3$ & $10^2\times k_4$ & $10^3\times k_5$ \\
                   &    [\um]   &  [\%] &  [\%] &  [\%] &  [\%] &  [\%] & $\lsun/\msun$ & $\lsun/\msun$ & $\lsun/\msun$  & $\lsun/\msun$    &  $\lsun/\msun$ }
\startdata
M51a     & 24&$35.0^{+14.0}_{-10.7}$&$15.5^{+7.4}_{-5.8}$&$\bf41.0^{+10.8}_{-11.5}$&$ 5.9^{+3.8}_{-3.6}$&$ 2.6^{+2.7}_{-1.8}$&  4.7$^{+  0.5}_{-  0.4}$& 2.4$^{+ 0.3}_{- 0.2}$& 2.0$^{+0.2}_{-0.2}$& 1.6$^{+0.8}_{-0.8}$&  5.4$^{+ 4.9}_{- 3.7}$\vspace{0.06in}\\
M51a     & 70&$33.9^{+12.3}_{-10.6}$&$15.3^{+7.8}_{-5.3}$&$\bf37.0^{+10.3}_{-10.6}$&$ 8.6^{+4.4}_{-4.3}$&$ 5.3^{+3.0}_{-2.7}$& 22.6$^{+  2.8}_{-  3.1}$&12.3$^{+ 1.6}_{- 1.5}$& 9.2$^{+1.2}_{-1.3}$&12.2$^{+3.1}_{-4.1}$& 51.9$^{+23.2}_{-15.9}$\vspace{0.06in}\\
M51a     &160&$26.0^{+ 8.2}_{- 7.1}$&$ 9.8^{+6.8}_{-3.8}$&$\bf37.5^{+11.6}_{-10.5}$&$17.2^{+5.6}_{-5.9}$&$ 9.5^{+5.1}_{-5.4}$& 17.5$^{+  2.3}_{-  2.1}$& 7.9$^{+ 1.5}_{- 1.4}$& 9.5$^{+1.0}_{-1.2}$&23.7$^{+3.4}_{-2.6}$& 90.6$^{+12.8}_{-15.3}$\vspace{0.06in}\\
M51a     &250&$25.9^{+ 7.5}_{- 7.1}$&$ 7.7^{+6.8}_{-3.1}$&$\bf37.5^{+12.8}_{-10.8}$&$19.1^{+6.1}_{-5.8}$&$ 9.8^{+6.6}_{-5.9}$&  4.7$^{+  1.2}_{-  0.7}$& 1.8$^{+ 0.5}_{- 0.5}$& 2.7$^{+0.5}_{-0.5}$& 7.5$^{+1.5}_{-1.2}$& 25.8$^{+ 7.6}_{- 6.7}$\vspace{0.06in}\\
M51a     &TIR&$30.2^{+11.5}_{- 8.6}$&$12.7^{+7.4}_{-4.8}$&$\bf38.3^{+10.7}_{-11.0}$&$12.0^{+4.3}_{-4.6}$&$ 6.8^{+3.2}_{-3.4}$& 45.2$^{+  8.1}_{-  6.8}$&22.1$^{+ 4.3}_{- 3.1}$&20.7$^{+4.0}_{-3.6}$&35.5$^{+9.0}_{-7.5}$&134.7$^{+38.9}_{-29.3}$\vspace{0.06in}\\\hline\\
M51b     & 24&$18.8^{+ 8.3}_{- 6.8}$&$12.0^{+6.4}_{-4.5}$&$ 7.9^{+ 4.4}_{- 3.9}$&$15.7^{+4.8}_{-5.0}$&$\bf45.6^{+5.1}_{-5.7}$&  4.5$^{+  0.6}_{-  0.6}$& 9.8$^{+ 4.3}_{- 3.5}$& 2.8$^{+0.6}_{-0.7}$& 1.4$^{+0.3}_{-0.4}$& 12.8$^{+ 2.2}_{- 1.6}$\vspace{0.06in}\\
M51b     & 70&$24.2^{+ 9.2}_{- 8.9}$&$ 5.0^{+1.9}_{-1.4}$&$ 3.4^{+ 3.6}_{- 1.9}$&$18.8^{+6.4}_{-6.9}$&$\bf48.6^{+4.3}_{-6.8}$& 27.6$^{+  3.5}_{-  3.9}$&20.1$^{+ 6.2}_{- 5.5}$& 6.6$^{+2.6}_{-2.7}$& 7.5$^{+2.8}_{-2.4}$& 65.5$^{+ 9.0}_{- 9.0}$\vspace{0.06in}\\
M51b     &160&$23.5^{+ 8.1}_{- 6.5}$&$10.8^{+4.3}_{-3.5}$&$ 7.8^{+ 4.1}_{- 3.8}$&$22.3^{+4.1}_{-3.8}$&$\bf35.6^{+3.8}_{-5.6}$& 16.8$^{+  2.4}_{-  2.5}$&25.2$^{+10.0}_{- 6.0}$& 8.1$^{+2.0}_{-1.5}$& 5.7$^{+0.9}_{-1.0}$& 29.2$^{+ 4.3}_{- 4.0}$\vspace{0.06in}\\
M51b     &250&$23.1^{+ 7.2}_{- 5.4}$&$14.1^{+4.6}_{-4.3}$&$ 9.5^{+ 4.4}_{- 4.3}$&$\bf27.2^{+3.2}_{-3.8}$&$26.2^{+3.6}_{-4.0}$&  4.7$^{+  1.3}_{-  1.1}$& 9.4$^{+ 3.3}_{- 2.9}$& 2.8$^{+0.9}_{-0.8}$& 2.0$^{+0.3}_{-0.4}$&  6.1$^{+ 1.3}_{- 1.1}$\vspace{0.06in}\\
M51b     &TIR&$22.7^{+ 8.9}_{- 7.9}$&$ 7.8^{+3.1}_{-2.5}$&$ 5.9^{+ 4.1}_{- 2.9}$&$17.9^{+4.8}_{-5.2}$&$\bf45.8^{+3.9}_{-6.7}$& 47.2$^{+  7.7}_{-  7.8}$&56.1$^{+20.8}_{-17.6}$&18.6$^{+5.4}_{-5.4}$&13.0$^{+4.0}_{-3.3}$&109.7$^{+22.8}_{-15.9}$\vspace{0.06in}\\\hline\\
outskirts& 24&$\bf40.1^{+ 9.4}_{- 8.2}$&$19.3^{+5.9}_{-5.2}$&$39.1^{+ 8.6}_{- 8.5}$&$ 1.4^{+1.5}_{-1.0}$&$ 0.3^{+0.7}_{-0.3}$&  3.3$^{+  0.3}_{-  0.3}$& 1.9$^{+ 0.2}_{- 0.2}$& 1.3$^{+0.1}_{-0.1}$& 0.2$^{+0.2}_{-0.1}$&  1.0$^{+ 2.6}_{- 1.0}$\vspace{0.06in}\\
outskirts& 70&$\bf42.4^{+10.0}_{- 8.0}$&$19.5^{+6.2}_{-5.1}$&$38.1^{+ 8.5}_{- 7.9}$&      \nodata       &      \nodata       & 17.2$^{+  2.0}_{-  2.3}$& 9.1$^{+ 1.3}_{- 1.0}$& 6.3$^{+0.9}_{-0.9}$&      \nodata       &       \nodata         \vspace{0.06in}\\
outskirts&160&$\bf41.5^{+ 8.4}_{- 8.1}$&$17.4^{+5.8}_{-4.2}$&$36.2^{+ 9.3}_{- 8.5}$&$ 4.1^{+1.3}_{-1.1}$&$ 0.8^{+0.7}_{-0.6}$& 23.7$^{+  2.9}_{-  3.1}$&12.2$^{+ 1.6}_{- 1.6}$& 8.5$^{+1.2}_{-1.0}$& 3.7$^{+0.7}_{-0.7}$& 19.7$^{+11.9}_{-10.6}$\vspace{0.06in}\\
outskirts&250&$\bf40.5^{+ 7.5}_{- 7.9}$&$14.0^{+5.0}_{-3.4}$&$34.9^{+ 8.8}_{- 8.6}$&$ 8.6^{+2.1}_{-2.1}$&$ 2.0^{+1.6}_{-1.3}$&  7.5$^{+  1.6}_{-  1.2}$& 3.2$^{+ 0.7}_{- 0.5}$& 2.7$^{+0.5}_{-0.5}$& 2.5$^{+0.5}_{-0.4}$& 15.5$^{+ 6.6}_{- 5.9}$\vspace{0.06in}\\
outskirts&TIR&$\bf41.7^{+ 9.6}_{- 7.4}$&$18.3^{+6.1}_{-4.5}$&$37.4^{+ 8.8}_{- 8.9}$&$ 2.6^{+1.4}_{-0.9}$&$ 0.0^{+0.5}_{-0.0}$& 44.9$^{+  8.3}_{-  7.0}$&24.1$^{+ 3.7}_{- 4.0}$&16.4$^{+3.3}_{-2.5}$& 4.5$^{+1.7}_{-1.2}$&  2.9$^{+24.2}_{- 2.9}$\vspace{0.06in}\\\hline\\
all      & 24&$34.6^{+13.7}_{- 9.6}$&$16.2^{+7.2}_{-6.0}$&$\bf40.6^{+12.1}_{-12.2}$&$ 2.5^{+1.3}_{-1.2}$&$ 6.1^{+1.6}_{-1.4}$&  4.6$^{+  0.5}_{-  0.4}$& 2.6$^{+ 0.4}_{- 0.3}$& 2.1$^{+0.3}_{-0.2}$& 0.5$^{+0.2}_{-0.2}$&  7.1$^{+ 0.9}_{- 0.9}$\vspace{0.06in}\\
all      & 70&$36.9^{+12.5}_{- 9.3}$&$15.2^{+8.1}_{-5.3}$&$\bf37.3^{+11.5}_{-10.7}$&$ 4.1^{+1.8}_{-1.4}$&$ 6.5^{+1.5}_{-1.4}$& 24.1$^{+  3.1}_{-  2.8}$&12.6$^{+ 1.9}_{- 1.7}$& 9.7$^{+1.4}_{-1.7}$& 4.2$^{+1.3}_{-1.5}$& 37.5$^{+ 5.6}_{- 4.7}$\vspace{0.06in}\\
all      &160&$37.7^{+12.2}_{- 9.7}$&$14.1^{+8.5}_{-5.0}$&$\bf38.5^{+11.8}_{-11.7}$&$ 4.8^{+1.5}_{-1.2}$&$ 4.9^{+1.3}_{-1.0}$& 24.7$^{+  3.1}_{-  3.3}$&11.7$^{+ 2.2}_{- 1.9}$&10.0$^{+1.4}_{-1.5}$& 4.7$^{+0.9}_{-0.8}$& 28.3$^{+ 3.9}_{- 3.9}$\vspace{0.06in}\\
all      &250&$38.8^{+11.4}_{- 9.1}$&$11.6^{+8.2}_{-4.5}$&$\bf38.8^{+12.7}_{-12.0}$&$ 7.1^{+2.3}_{-1.8}$&$ 3.6^{+1.2}_{-1.0}$&  7.2$^{+  1.6}_{-  1.2}$& 2.8$^{+ 0.7}_{- 0.7}$& 2.9$^{+0.5}_{-0.5}$& 2.1$^{+0.4}_{-0.4}$&  5.9$^{+ 1.5}_{- 1.4}$\vspace{0.06in}\\
all      &TIR&$36.9^{+12.1}_{- 9.8}$&$14.7^{+8.2}_{-5.2}$&$\bf38.0^{+12.1}_{-11.4}$&$ 3.9^{+1.3}_{-1.1}$&$ 6.6^{+1.6}_{-1.4}$& 53.2$^{+ 10.1}_{-  8.8}$&26.4$^{+ 5.2}_{- 4.0}$&21.7$^{+4.2}_{-3.9}$& 8.6$^{+2.4}_{-2.2}$& 82.3$^{+18.2}_{-11.6}$\vspace{0.06in}
\enddata
\label{tab:ir_fractions}
\end{deluxetable*}

In reality, we are able to derive the age-dependent light-to-mass ratios through this procedure for each SFH component. The last columns of Table~\ref{tab:ir_fractions} shows, for each region, the derived IR luminosity at any given band over the stellar mass, separately for each SFH step component. In Figure~\ref{fig:lm_all}, we display the IR luminosities over current stellar mass associated to each of the five SFH steps. The figure shows the result of decomposing the IR bands using all pixels (as in the last rows of Table~\ref{tab:ir_fractions}). This way, we can actually see that dust in different environments, heated by stars of the same age range, generally have the same luminosity per stellar mass at any region. Some of these $k_i$ values, however, are changing from region to region and sometimes are significantly different in the outskirt region. We plan to investigate this variation in more detail in a future paper, with a larger sample of spatially-resolved galaxies.

We find old stars do heat dust much less efficiently than young stars (per stellar mass), as shown in  Table ~\ref{tab:ir_fractions} and Figure~\ref{fig:lm_all}. Even though the ratio of IR luminosity over stellar mass does drop consistently with age, old stars may outshine younger stars at any of the IR wavelengths studied here, including the TIR luminosity. In fact they do in M51b.

\begin{figure*}
  \singlespace
  \includegraphics[width=\textwidth]{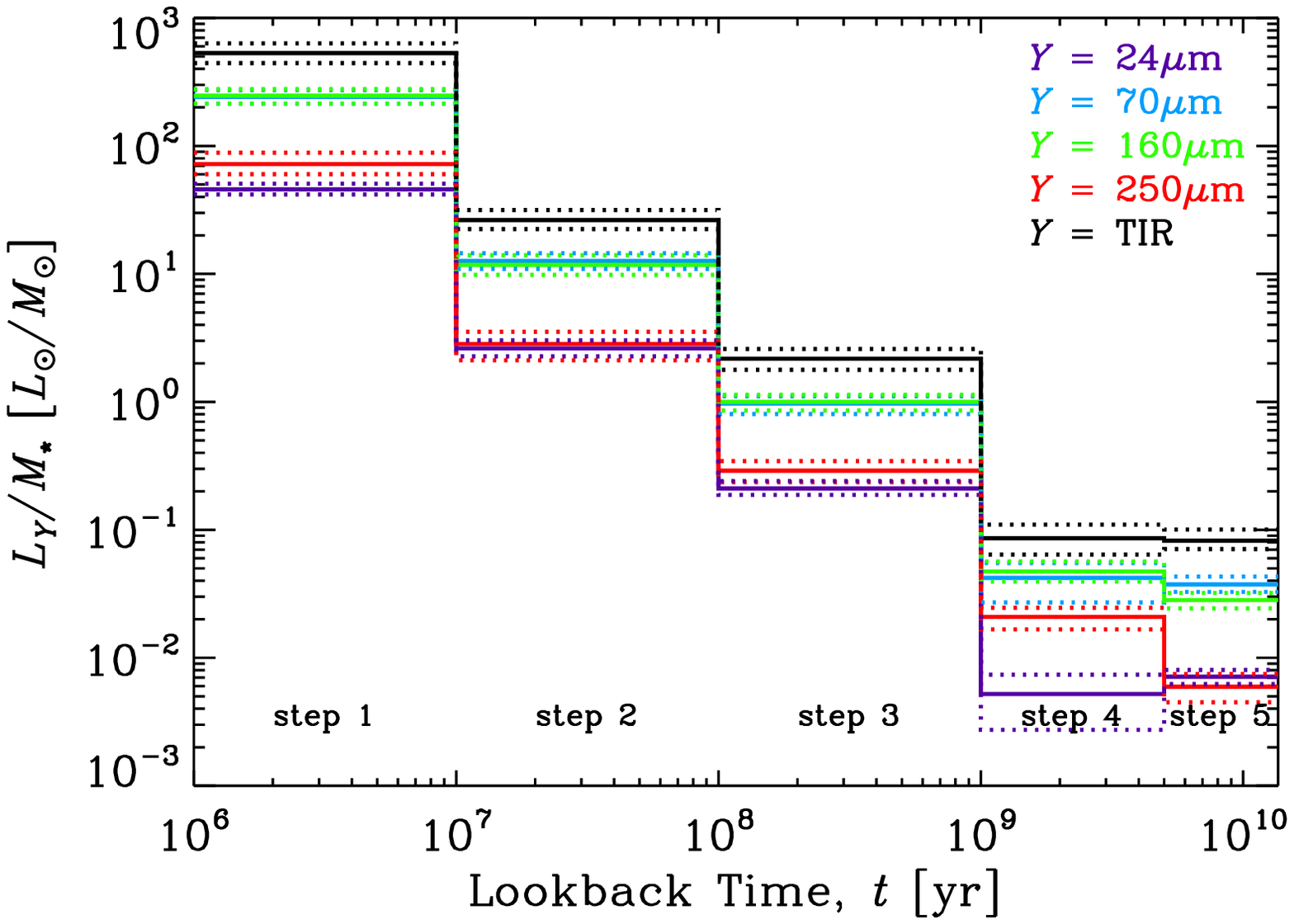}
  \vspace{-0.2in}
  \caption{Age-dependent light-to-mass for all FIR bands and the TIR associated to all our five SFH steps, as estimated from the whole map of M51. The first SFH step is actually from 0 to 10~Myr, but it is displayed here from 1 to 10~Myr for simplicity. Violet, cyan, green, red, and black respectively represent the 24, 70, 160, 250, and TIR luminosities divided by the stellar mass of each SFH step. Dotted lines show the 16th and 84th percentiles for each quantity. The ratio of IR luminosity over stellar mass drops consistently with age, meaning the old stars do heat dust much less efficiently than young stars (per stellar mass). Nevertheless, old stars may, and in fact they do in M51b, outshine younger stars at any of the IR wavelengths studied here, including the TIR luminosity (see Table~\ref{tab:ir_fractions}).
}
  \label{fig:lm_all}
\end{figure*}

\section{Summary}\label{sec:summary}

We presented a new broadband SED fitting procedure to efficiently derive star-formation histories for a large number of SEDs. We used the Whirlpool system, M51 (composed of the galaxies M51a and M51b) as a initial testbed for our code, {\sc Lightning}, taking advantage of the fact that extensive data and previous studies are available for M51. For this work, we made use of FUV--to--FIR data, including 18 broadband images of M51 from \galex, SDSS, 2MASS, \spitzer, and \herschel.

We developed a method based solely on \spitzer\ bands to detect foreground stars in the Galaxy, masked the brightest foreground stars, and replaced these pixels with the local background in all images from the FUV to \spitzer\ 4.5\,\um. We then convolved all images to a common spatial resolution of 25\arcsec\ FWHM and registered them to a common astrometric frame with 10\arcsec\ pixels. At the distance of 8.58\,Mpc, the pixel size corresponds to 416\,pc. The FIR images (24, 70, 160, and 250\,\um) were combined to produce a TIR luminosity map, based on \cite{Galametz2013} calibration of \cite{DL2007} models.

We employed {\sc Lightning} to individually fit 2043 pixels, each with 12 photometric bands from \galex\ FUV to \spitzer\ 4.5\,um and a TIR luminosity. We model the SFH of each pixel as a function of time with five steps: 0--10\,Myr, 10--100\,Myr, 0.1,1\,Gyr, 1--5\,Gyr, and 5--13.6\,Gyr. The total attenuated power is constrained by the observed TIR and we model the extinction curve with three parameters. The code has a 3D parameter grid for the extinction parameters and maximizes likelihood in each position of parameter space by an efficient inversion procedure that uniquely determines all five SFH step intensities at once. We have derived 400 extra simulated maps to properly derive uncertainties for all quantities. We do not employ any dust emission modeling in this version of the code, but we plan to add this feature in a later version. This way we have produced reliable SFH and extinction maps for M51, which we make publicly available to the astronomy community at {\ttfamily https://lehmer.uark.edu/}. We have compared the TIR luminosities, stellar masses, and overall SFHs on a pixel-by-pixel basis with the publicly available code CIGALE and found remarkable agreement.

We also tested the hybrid SFR tracer combining UV and IR wavelengths and observed similar trends to the ones previously found by \cite{Eufrasio2014} and \cite{Boquien2016}. The correction factor needed to convert a given IR band intensity into the attenuated UV luminosity was found to increase for increasing sSFR100, i.e., younger regions show more efficient UV absorption per unit IR emission. This seems to imply the older stellar population, not associated with the recent SFR of the last 100\,Myr, is still able to considerably heat dust, many times outshining younger regions in the IR.

With this in mind, we employed a procedure to determine the fraction of luminosity associated with each SFH step at each FIR wavelength. Interestingly, we found significant emission from all five SFH steps, at all FIR wavelengths. In fact, dust heated by the oldest stellar population (5--13.6\,Gyr) outshines each of the other four in M51b. In M51a, the IR emission mainly arises from dust heated by the population most likely produced with the interaction (third SFH step, 0.1--1\,Gyr). And the outskirts of the interacting system emits almost equal amounts of IR emission associated with the youngest (0--10\,Myr) SFH step and well as the third (0.1--1\,Gyr). From this decomposition, we were also able to derive the ratio of IR luminosity at a given band and stellar mass associated with each SFH step, for the four large regions investigated here. The general trend is a decline of dust luminosity per stellar mass for all the wavelengths considered here (24, 70, 160, and 250\,\um\ and the TIR). In other words, we find stars of all ages heat dust. Also, younger the stellar population, more efficient this heating. But surprisingly, older stars may outshine the youngest stellar population even in a spiral galaxy, depending on the SFH of the galaxy.

Our work shows how a simple SFH model as steps of time can be useful and opens a promising possibility of decomposing any given map of a galaxy (be it X-rays, IR, or radio) into various age bins to determine its time-evolution. In fact, we have already successfully decomposed the X-ray binary luminosity functions across M51 into the five steps used here to empirically determine its evolution. This will be published under a separately paper (Lehmer et al. 2017, submitted).

\acknowledgments 
We thank the anonymous referee for comments that significantly improved the paper.
We gratefully acknowledge support from NASA/ADAP grant NNX13AI48G (B.D.L., R.T.E., A.Z.). R.T.E. and E.D. acknowledge NASA ADAP proposal NNH11ZDA001N. A.Z. acknowledges funding from the European Union's Seventh Framework Programme (FP/2007--2013)/ERC Grant Agreement n. 617001.
Based on observations made with the NASA {\it Galaxy Evolution Explorer}. \galex\ is operated for NASA by the California Institute of Technology under NASA contract NAS5-98034.
This publication makes use of data products from the Two Micron All Sky Survey, which is a joint project of the University of Massachusetts and the Infrared Processing and Analysis Center/California Institute of Technology, funded by the National Aeronautics and Space Administration and the National Science Foundation.
This work is based on observations made with the Spitzer Space Telescope, obtained from the NASA/IPAC Infrared Science Archive, both of which are operated by the Jet Propulsion Laboratory, California Institute of Technology under a contract with the National Aeronautics and Space Administration.


{\it Facilities:}
\facility{\galex}, 
\facility{Sloan},
\facility{FLWO:2MASS},
\facility{\spitzer},
\facility{\herschel}



\end{document}